\documentclass[twocolumn, switch]{article}
\usepackage[left=1.5cm, right=1.5cm, top=1.5cm, bottom=1.5cm]{geometry}
\usepackage{cite}
\usepackage{cuted}
\usepackage{abstract}
\usepackage{amsmath,amssymb,amsfonts}
\usepackage[dvipdfm]{graphicx} \usepackage{bmpsize}
\usepackage{algorithmic}
\usepackage{graphicx,color}
\usepackage{textcomp}
\usepackage{makecell}
\usepackage{graphicx}
\usepackage{placeins}
\usepackage{float}
\usepackage{subcaption}
\usepackage[long]{optidef}
\usepackage{cite}
\usepackage{xcolor}
\usepackage{subfig}
\usepackage{amssymb}
\usepackage{color}
\usepackage{multirow}
\usepackage{tabularx}
\usepackage{array}
\usepackage{multirow}
\usepackage{threeparttable}
\usepackage{lipsum}
\usepackage{booktabs}
\usepackage{longtable}
\usepackage{enumitem}
\usepackage{balance}

\begin{document}
\title{Deep Learning-based Techniques for Integrated Sensing and Communication Systems: State-of-the-Art, Challenges, and Opportunities }

\author{Murat Temiz\footnote{Department of Electronic and Electrical Engineering, Faculty of Engineering Sciences, University College London, WC1E 7JE, London, United Kingdom.}~\footnote{Department of  Electrical and Electronics Engineering, Faculty of Engineering, Middle East Technical University, 06800, Ankara, Turkey.},
        Yongwei Zhang\footnote{School of Transportation and Civil Engineering, Nantong University, Nantong, JS, 226019, China.}, 
        Yanwei Fu$^\ddagger$,
        Chi Zhang$^\ddagger$,
        Chenfeng Meng$^\ddagger$,
        \\ Orhan Kaplan\footnote{Department of  Electrical and Electronics Engineering, Faculty of Technology, Gazi University, 06560, Ankara, Turkey.},
        Christos Masouros$^*$}

\date{}
\maketitle
\begin{strip}
    \centering
    \begin{minipage}{1\textwidth}
        \begin{abstract}
            This article comprehensively reviews recent developments and research on deep learning-based (DL-based) techniques for integrated sensing and communication (ISAC) systems. ISAC, which combines sensing and communication functionalities, is regarded as a key enabler for 6G and beyond networks, as many emerging applications, such as vehicular networks and industrial robotics, necessitate both sensing and communication capabilities for effective operation. A unified platform that provides both functions can reduce hardware complexity, alleviate frequency spectrum congestion, and improve energy efficiency. However, integrating these functionalities on the same hardware requires highly optimized signal processing and system design, introducing significant computational complexity when relying on conventional iterative or optimization-based techniques. As an alternative to conventional techniques, DL-based techniques offer efficient and near-optimal solutions with reduced computational complexity. Hence, such techniques are well-suited for operating under limited computational resources and low latency requirements in real-time systems. DL-based techniques can swiftly and effectively yield near-optimal solutions for a wide range of sophisticated ISAC-related tasks, including waveform design, channel estimation, sensing signal processing, data demodulation, and interference mitigation. Therefore, motivated by these advantages, recent studies have proposed various DL-based approaches for ISAC system design. After briefly introducing DL architectures and ISAC fundamentals, this survey presents a comprehensive and categorized review of state-of-the-art DL-based techniques for ISAC, highlights their key advantages and major challenges, and outlines potential directions for future research.
        \end{abstract}
        \hspace{2cm}
    \end{minipage}
\end{strip}

\section{Introduction}

Integrated sensing and communication (ISAC) has emerged as a key technology for future-generation wireless networks such as 6G and beyond. Delivering both radar sensing and wireless communication functions using the same waveform and hardware reduces radio spectrum congestion and the amount of hardware used and improves the energy efficiency of the networks. ISAC systems can utilize various waveforms to provide both sensing and communication, depending on the primary function of the system. For instance, communication waveforms can be optimized to enable some degree of sensing capabilities without significantly reducing communication performance. This will enable the communication network and devices to perform sensing while communicating with each other. On the other hand, sensing waveforms can also be employed to provide communication by embedding communication data in these signals. 

While ISAC provides significant advantages over current sensing and communication networks, the design and implementation of ISAC systems can be challenging, especially in terms of signal processing and computational complexity. Because combining both systems requires solving optimization problems for generating the optimum waveform or processing the received signals for sensing and communication at the receiver, considering the limited computational resources of hardware and the strict timing constraint imposed by the varying channel conditions, it can be challenging to reach optimum solutions \cite{AshrafJointOptimization2023, temiz2021optimized} via conventional signal processing techniques. This high complexity is especially encountered in digital signal processing and optimization stages while generating the optimum waveform at the transmitter and processing the received signals for communications and sensing at the receivers.  

The increased complexity of the ISAC design requires new approaches instead of the ones based on mathematical optimization or iterative techniques. Machine learning-based (ML-based) techniques, especially deep learning-based (DL-based) techniques, can be leveraged for ISAC systems to mitigate the complexity and address these challenges, enabling their seamless integration within future communication networks. Thus, DL-based techniques have recently been proposed to achieve near-optimum solutions for these challenges and issues in ISAC systems with limited hardware and energy resources. 

Recent surveys have systematically reviewed the ISAC studies from different perspectives \cite{Zhangjoint2021, LiuSurveyisac2022,  ZhouSurveyISACWaveform2022, Wei6GISACSurvey2023, WangSurveyISAC2022, LiuISAC_RIS2023, LuISACRecentChallenges2024}. Liu et al. aimed to investigate the fundamental limits of ISAC techniques from a theoretical perspective \cite{LiuSurveyisac2022}, while Zhou et al. explored waveform design techniques and studies for ISAC \cite{ZhouSurveyISACWaveform2022}. Furthermore, the promising solutions and benefits that would be brought by ISAC and intelligent reflective surfaces (IRS) to 6G networks are surveyed in \cite{Wei6GISACSurvey2023, LiuISAC_RIS2023}, and enabling techniques, future applications, data sets, and future directions are reviewed in \cite{WangSurveyISAC2022}. Finally, recent advancements and open challenges in ISAC systems are also explored and presented \cite{LuISACRecentChallenges2024}. All of these surveys have focused on various aspects, problems, and solutions for ISAC systems. {Moreover, a review article has focused on machine learning (ML) enhanced sensing and ISAC signal design \cite{AdeSurvey2024}, and an earlier study identified ten potential roles of ML in ISAC systems \cite{demirhan2022integrated}.} 

On the other hand, ML and deep learning (DL) have been extensively investigated to improve {the performance of communication networks}. Such studies focusing on ML, {DL}, {reinforcement learning (RL)}, and distributed learning for future communication networks, including Internet of Things (IoT) systems, 6G networks, and vehicular networks, are also reviewed \cite{JaganMLIoT2019, DuML6G2020,LuongDLCom2019, ZhangDLCom2019, HuDistributedMLCom2021, AhmadMLComm2020}. Especially {DL} and RL are promising techniques for physical layer problems \cite{DörnerDLComm2018}, while distributed learning is seen as an essential approach for networking problems \cite{liu2022distributed}. {By integrating DL-based techniques, ISAC systems can become more intelligent, flexible{,} and efficient, paving the way for advanced applications in various domains, such as 6G networks or next-generation sensing networks.} Table~\ref{table:overview}\footnote{Note: Some publications are categorized under more than one category in Table~\ref{table:overview} since they propose DL-based architectures for multiple functions such as jointly demodulating communication data and sensing processing} presents the summary of the DL-based studies and categorizes them according to their applications and algorithms used. 

\begin{table*}[]
\caption{List of Surveys on ISAC Study and Design}
\label{tab:surveys}
\centering
\small
\begin{tabular}{|l|l|}
\hline
\textbf{Survey} & \textbf{Topic} 
\\  \hline
 \cite{Zhangjoint2021} 2021  & Overview of signal processing techniques for ISAC systems.
\\  \hline
 \cite{LiuSurveyisac2022} 2022  &  Fundamental limits of ISAC techniques from a theoretical \\ & perspective.      
  \\ \hline
\cite{ZhouSurveyISACWaveform2022}   2022    &    Waveform design techniques and studies for ISAC.    
\\ \hline
\cite{WangSurveyISAC2022} 2022 &  Enabling techniques for ISAC, applications, tools \\ & and data sets, standardization, and future directions \\ & in ISAC research. \\ \hline
\cite{Wei6GISACSurvey2023} 2023 & ISAC techniques for 5GA and 6G networks \\ \hline
\cite{LiuISAC_RIS2023} 2023 & ISAC systems powered by reflective intelligent surfaces. \\ \hline
\cite{LuISACRecentChallenges2024} 2024      & Recent advancements and open challenges in ISAC.  \\ \hline
\cite{AdeSurvey2024} 2024      & {Machine
learning enhanced ISAC systems.}  \\ \hline
\end{tabular}
\end{table*}
\begin{table}
\caption{Acronyms used in this article.}
    \centering
    \small
    \begin{tabular}{|c|l|}
    \hline
AI & Artificial Intelligence \\\hline
{BER} & {Bit Error Rate} \\ \hline
{BLER} & {Block Error Rate} \\ \hline
CNN & Convolutions neural network \\ \hline
CRLB & Cramer-Rao lower bound \\ \hline
CSI & Channel state information \\ \hline
DFRC & Dual-Functional Radar-Communication \\ \hline
NN  & Neural network  \\ \hline
DNN  & Deep neural network  \\ \hline
DL  & Deep learning  \\ \hline
DRL & Deep reinforcement learning \\\hline
ELM & Extreme learning machine \\\hline
FIM  & Fisher information matrix \\ \hline
FCDNN & Fully connected deep neural Network \\ \hline
{FL} & {Federated learning} \\ \hline
FMCW & Frequency-modulated continuous-wave \\\hline
FNN & Feed-forward neural network \\\hline
{GAN} & {Generative adversarial network} \\\hline
RL & Reinforcement learning \\\hline
IM    & Index modulation \\ \hline
ISAC    & Integrated sensing and communications \\ \hline
{IRS}   & {Intelligent reflective surface} \\ \hline
IoT & Internet of Things\\ \hline
KNN&K-nearest neighbour\\ \hline
LSTM&Long short-term memory\\ \hline
{LMMSE} & {Linear minimum mean-square error} \\ \hline
MFOL & Model-free online learning\\ \hline
MBOL & Model-based online learning \\ \hline
ML&Machine learning\\\hline
MLP & Multilayer perceptron \\ \hline
MSE & Mean squared error \\ \hline
NMSE & Normalized mean squared error \\ \hline
NOMA & Non-orthogonal multiple access \\ \hline
{OCDM} & {Orthogonal chirp-division mult.} \\ \hline
OFDM & Orthogonal frequency-division mult. \\ \hline
OTFS & Orthogonal time frequency space \\ \hline
{QNN} & {Quantum neural network} \\ \hline
ReLU & Rectified linear unit activation function \\ \hline
{RMSE} & {Root mean squared error} \\ \hline
SER & Symbol error rate \\ \hline
SLP & Symbol-level precoder \\ \hline
SVM & Support vector machine\\ \hline
SNN & Spiking neural network \\ \hline
MUSIC & Multiple signal classification\\ \hline
THz & Terahertz\\ \hline
UE & User equipment\\ \hline
MIMO & Multiple-input multiple-output \\ \hline
    \end{tabular}
    \label{tab:acry}
\end{table}

 

\subsection*{Contributions}
{In contrast to the aforementioned surveys on ISAC systems, this article explores state-of-the-art advancements and research on DL-based techniques for ISAC systems and networks and presents a list of challenges, open research questions, and research opportunities. This study categorically and systematically reviews DL-based techniques developed for various modules of the ISAC systems, such as for beamforming, channel estimation, demodulation, and sensing processing. Afterward, it presents the advantages of DL-based techniques in terms of performance enhancements and computational complexity reduction in comparison with conventional methods. Moreover, it also highlights challenges encountered in DL-based ISAC techniques and possible future research directions in the intersection of DL-based techniques and ISAC systems for the development of future-generation sensing and communication networks.}

The contributions of this study are summarized as follows:
\begin{itemize}
    \item {This study categorically and systematically reviews the DL-based techniques developed for the ISAC system design and optimization, driven by a need to achieve real-time} implementation, reduced system complexity, and improved performance and energy efficiency.
    \item {It also explores widely employed DL architectures, and their suitable applications for various modules of ISAC systems, and examines their computational complexities for training and inference, and dataset size requirements.} 
    \item {It identifies the challenges, open problems, and future research directions in DL-based techniques and their implementation for ISAC systems and networks.}
\end{itemize}

The rest of this article is organized as follows. Section II describes the background of {DL} and provides examples of its applications in communication and sensing systems. Section III reviews the ISAC fundamentals and various architectural challenges, solutions, and performance metrics. Section IV presents the DL-based techniques for transmitter design and waveform optimization. Section V describes {DL} methods for channel estimation. Section VI presents DL-based techniques for communication and sensing receiver design. {Section VII examines the computational complexity of DL-based techniques in comparison with conventional optimization-based techniques.} {Open research questions and} potential future research directions are outlined in Section VIII. Finally, the conclusion of this article is drawn in Section IX. {For convenience, all acronyms used in this article are given in Table~\ref{tab:acry}.}

\begin{table*}
\caption{Overview of DL-based ISAC techniques. }\label{table:overview}
\label{tab:xxx}
\centering
\small
\begin{tabular} {|m{2cm}<{\centering}|m{2cm}<{\centering}|m{2cm}<{\centering}|m{9cm}<{\centering}|}

	\hline
	\textbf{Application Category}& \textbf{Algorithms} & \textbf{References} & \textbf{Key features} \\
	\hline 
Waveform Design and Optimization & DNN, RL, LSTM, FCDNN  &   \cite{PulkinenISAC2024}, \cite{PulkkinenMFOL2023}, \cite{zheng2024endtoend}, \cite{RL2023},\cite{AE2022}, \cite{XuRecon2024}\cite{liu2023distributed}\cite{TemizUnsupervised2025}& 

\begin{itemize}
    \item ISAC Waveform design and power allocation,
    \item Waveform optimization for dynamic environmental conditions.
    \item Optimum waveform for sensing and communications
    \item Optimization of peak-to-average power ratio
    \item Codebook design, signal modulation, power allocation, dynamic waveform design, energy-efficient algorithms, optimization
\end{itemize}
\\ \hline

Predictive Beamforming & DNN, CNN, LSTM, DRL, RL & \cite{beamforming2022},\cite{LiuCLSTM2022},\cite{LiuPreBeam2022},\cite{ZhangTransformerPred2024},\cite{wang2023intelligent},\cite{ZhangIntegratedProactive2024},\cite{zhang2024predictive},\cite{Mubeamforming2021}& 

\begin{itemize}
    \item Beamforming and precoder design for ISAC systems.
    \item Predictive beamforming design
    \item Interference management and exploitation
\end{itemize}
 \\  \hline

Channel ~~~~~ Estimation & FCDNN, CNN, ELM&  \cite{LiuCSI2022, LiuDLCSI2023}, \cite{CSI-DRSN-mcf}, \cite{LiuEL_CSIA2023}& 

\begin{itemize}
    \item Channel estimation for ISAC systems.
    \item Angle of arrival estimation.
    \item Channel tracking and prediction.  
\end{itemize}

\\ \hline

Communication Receiver & FCDNN, SNN, LSTM, DNN Transformer & \cite{wuSensingIntegratedDFTSpread2023},
\cite{chenNeuromorphicIntegratedSensing2023a}, \cite{liuDeepLearningBased2022}, \cite{hu2024isac}, \cite{jiang2024isac}& 

\begin{itemize}
    \item Communication data demodulation.
    \item Interference and self-interference cancellation and exploitation.
    \item UWB impulse communication receiver
\end{itemize}
\\ \hline

Sensing Receiver & FCDNN, DNN, DNN Transformer, CNN, QNN,  &\cite{wuSensingIntegratedDFTSpread2023},
\cite{chenNeuromorphicIntegratedSensing2023a},
\cite{hu2024isac},
\cite{jiang2024isac},\cite{gao5GNRHighPrecision2022},
\cite{KoikeQNN2022},
\cite{suarezDeepLearningaidedRobust2023},
\cite{liuVerticalFederatedEdge2022} &

\begin{itemize}
    \item Spectrum sensing for optimal resource allocation.
    \item Target detection, target classification, and recognition.
    \item Human gesture and motion recognition.
    \item Target range, velocity, angle estimation
    \item UE positioning
\end{itemize}
\\ \hline
\end{tabular}
\end{table*}

\section{Deep Learning Background}
\begin{figure}
    \centering \includegraphics[width=0.7\linewidth]{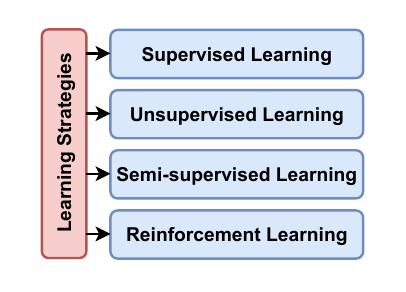}
    \caption{Learning strategies.}
    \label{fig:ML_strategies}
\end{figure}

{DL} techniques have been recently applied to many fields to optimize systems or solve complex problems that mathematical methods can not solve easily with limited computational resources and within a reasonably short time. {DL models can be trained via various learning strategies such as supervised, unsupervised, or {RL}. A summary of learning strategies is illustrated in Fig~\ref{fig:ML_strategies} and briefly explained below in this section.}

\subsection{Supervised Learning}
Supervised learning strategies utilize the labeled data sets during training and aim to create a model that maps the input data and the output data (labels) \cite{lecun2015deep}. This mapping can be highly sophisticated and complicated for large-scale problems, such that it cannot be easily explained mathematically. Supervised learning techniques generate models from this training data, and these models are used to classify the unlabelled data or create solutions for the problem \cite{cunningham2008supervised, suthaharan2016supervised}. {Various ML models, including support vector machines (SVM), logistic regression, deep neural networks (DNN), k-nearest neighbors (KNN), random forests, decision trees, boosted trees, can be trained via supervised learning \cite{jiang2020supervised,caruana2006empirical}.}

This learning type requires a large amount of labeled data for training. It strives to learn the relationship between the input and output data depending on the learning parameters during training. For instance, the mean squared error (MSE) can be used to construct the loss function for supervised learning, as given by
\begin{equation}
    \mathcal {L}_{s}(x,y,\theta_s) = \frac{1}{N}\sum _{(x_n,y_n)\in \mathbf{X}_{L}} (y_n - \hat{y}_n)^2,
\end{equation}
where $x_n$ and $y_n$ denote input and labeled data in the data set $\mathbf{X}_L$ consisting of N training samples, while $\theta_s$ and $\hat{y}_n$ denote the learning parameters and predicted output during the $n$th training instances, respectively. The loss function of supervised learning can also be defined based on various other metrics, such as cross-entropy, softmax loss function, and MSE, among other possible metrics \cite{Tian2022129}.

Requiring a large labeled dataset is the main disadvantage of this learning strategy, since obtaining a sufficiently large labeled dataset for most applications may not be possible. As a solution for the limited amount of available labeled data, various data augmentation methods are employed to increase the amount of labeled data by generating synthetic data in addition to the ground-truth data, especially for natural language processing and image processing applications \cite{shorten2021text, van2001art}. 

\subsection{Unsupervised Learning}
Unsupervised learning explores the similarities in finding patterns in the dataset without requiring labeled data. {This learning strategy is particularly effective for exploring relationships within the data, discovering underlying patterns, and solving optimization problems\cite{Ghahramani2004, NikbakhtUnsupervised2021}.}
 
More interestingly, especially for communications and sensing systems, {unsupervised learning can be performed to solve an optimization problem, where the loss function for training is defined by combining the objective function and constraints\cite{dizaji2018unsupervised, SunUnsupervised2019, NikbakhtUnsupervised2021, ye2024unsupervised}.} An exemplary loss function for unsupervised {DL} is given by
\begin{equation}
    \mathcal {L}_{u}(x,\theta_u) = -\sum_{k=1}^K \log_2 \left(1+\frac{f_k(x_n, \theta_u)}{f_i(x_n, \theta_u)+\sigma_n^2}\right),
\end{equation}
which aims to maximize the sum rate, i.e., minimizing the negative of the sum rate, where $\theta_u$ denotes the learning parameters of the unsupervised learning. Functions $f_k(.)$ and $f_i(.)$ compute the useful signal power and interference power on the $k$th user
equipment (UE), while $\sigma_n^2$ denotes the variance of the Additive white Gaussian noise at the $k$th UE. Moreover, it is also possible to consider other performance metrics, such as energy efficiency or probability of detection for sensing in the loss function of the unsupervised learning.

An unsupervised learning approach has been applied to resource allocation problems in wireless communication systems \cite{LiuOptimization2020} or beamformer and precoder design problems\footnote{{Across this article, \textit{beamformer} refers to the algorithms that directs the transmitted signals towards specific directions while \textit{precoder} refers to the algorithms that pre-process the signals before transmission by taking into account channel conditions to communicate with {multiple UEs}. However, \textit{beamformer} and \textit{precoder} terms are often used interchangeably in the literature.}} in ISAC systems \cite{ye2024unsupervised}. For ISAC, key performance parameters for sensing and communication {can be incorporated into the loss function of unsupervised learning.}

\subsection{Semi-supervised Learning}

Another approach in {DL} training is semi-supervised learning, where unlabeled data instances are utilized during learning in addition to the labeled data instances. {Especially in the case of not having a sufficient amount of labeled data, semi-supervised learning can achieve significantly higher accuracy compared to supervised learning \cite{van2020survey}. For instance, a semi-supervised learning-based radar data processing method outperforms the fully supervised learning-based method in object detection by up to 36\%, where the amount of labeled radar data is limited since labeling radar data is a difficult task \cite{LeeAutomotive2023}. Another study proposes a semi-supervised learning-based target parameter estimation method for ISAC systems, which can achieve similar performance with the supervised learning-based method by utilizing 98.8\% less labeled data\cite{MateosSemi2024}.} One of the approaches for semi-supervised learning is forming the loss function as a combination of the objective functions of supervised learning and unsupervised learning, which will yield a semi-supervised learning model as \cite{yang2022survey},  

\begin{equation} 
\mathcal{L} = \underset{\text{supervised loss}}{\underbrace{\alpha\sum _{(x,y)\in X_{L}}\mathcal {L}_{s}(x,y,\boldsymbol{\theta}_s)}}+\beta \underset{\text{unsupervised loss}}{\underbrace{\sum _{x\in X_{U}}\mathcal {L}_{u}(x,\boldsymbol{\theta}_u)}},
\end{equation}
where {hyperparameters $\alpha\in\mathbb{R}_{>0}$ and $\beta \in\mathbb{R}_{>0}$} are the weights of loss functions ($\mathcal{L}_s$ and $\mathcal{L}_u$) of supervised and unsupervised training, respectively. {The hyperparameters $\alpha$ and $\beta$ are generally fine-tuned during training, such that the importance of the unsupervised learning is gradually increased during training \cite{chen2022semi, laine2016temporal, oliver2018realistic}.} Moreover, $x$ and $y$ denote the input data and labeled data while $\boldsymbol{\theta}_s$ and $\boldsymbol{\theta}_u$ denote the learning parameters of supervised learning and unsupervised learning, respectively. 

\subsection{Reinforcement Learning}
RL is based on trial and error, where an agent strives to learn how to perform a task within an action space. The agent performs actions in the action space and observes the result of the actions and rewards to decide the subsequent actions. The main objective of the agent is to maximize the reward during the learning process. In this way, the agent can learn the tasks in a defined action space without having any labeled data \cite{kaelbling1996reinforcement, kuutti2020survey}. The main advantage of {RL} is that it does not require any labeled dataset. However, the learning process may require a large number of computationally intensive simulations since it employs a try-and-error approach to learning and requires a simulation environment.  

The objective of the {RL} problems is to choose actions {within the action space at each training instance} to maximize the rewards. For instance, the total reward received over $T$ time steps can be defined as the performance metric given by \cite{bartlett2003introduction},
\begin{align}
      J_T(i_0) & =  \mathbb{E}\left[\sum_{t=0}^T \alpha_d^t R(x_t)|_{x_0=i_0}  \right] \\ 
      & = \mathbb{E}\left[i_0 + \alpha_d R(x_1) + \alpha_d^2 R(x_2), \cdots, \alpha_d^N R(x_N)  \right] \notag,
\end{align}
where $i_0$, $R(.)$, and $0\leq \alpha_d <\leq 1$ denote the start state, reward function, and discount factor, respectively. When $\alpha_d$ is close to one, {RL} will pursue long-term rewards, while having a value closer to small will force it to strive for short-term rewards.

While {RL} can handle numerous tasks, it suffers from high-dimensional data sets, such as real-time images, video processing, or fusion of multiple sensor data. Accordingly, deep {RL (DRL)} was proposed to {handle} high-dimensional data sets by leveraging {DL} with {RL}\cite{arulkumaran2017deep}. DRL can effectively operate with real-time communication data, radar data, or visual data to learn {sophisticated tasks, as encountered in communication} networks \cite{FerianiDRL2021}, autonomous driving \cite{kiran2021deep}, and real-time control \cite{kuutti2020survey}.

\subsection{Transfer Learning}
Transfer learning pursues the reuse of the trained ML models for new problems, and it is one of the most powerful learning strategies considered for 6G and beyond networks \cite{girelli2023analysis} {since it can substantially reduce the training time}. {It can utilize} a {DL} model previously trained on a specific problem to solve a new problem that is unseen by the model. {The transfer learning approach has been used to reduce the amount of training data required for neural networks, optimize latency and energy efficiency, accelerate {RL} for communication network slicing, and design precoders for ISAC networks \cite{yang2021joint, nagib2023safe, coutinho2021transfer, liang2021transfer, liu2023distributed}.}

\subsection{Distributed {Learning}}

Training {large} {DL} models requires a large amount of data and computational power, exponentially increasing in relation to the number of parameters involved in the model. Furthermore, tuning the model's hyperparameters to devise the optimum {DL} model for a specific task may require re-training the model multiple times. Accordingly, numerous computational resources may need to be concurrently utilized to train comprehensive {DL} models, leading to a distributed computation network rather than a centralized system \cite{verbraeken2020survey}. 

Distributed learning may be performed by numerous strategies such as data parallelism, model parallelism, compute prioritization, and asynchronous parallel execution \cite{xing2016strategies}. However, sharing all the data or the entire model among all computation nodes may cause privacy concerns and a tremendous amount of communication overhead.  Thus, federated learning {(FL)} is devised as an efficient learning approach to overcome these problems. In {FL}, multiple nodes collaboratively train {DL} models without sharing the raw data, hence substantially reducing communication overheads among the nodes and also eliminating privacy concerns. Each node trains its own model with its own local data while only sharing the necessary parameters with an aggregation center and keeping its own raw data locally \cite{abdulrahman2020survey}. Owing to its advantages, {FL} has already been applied to {edge computing, IoT systems, vehicular communications, and ISAC systems \cite{chen2021distributed, liuVerticalFederatedEdge2022}.} 

\subsection{{Deep Learning Architectures}}

\begin{figure}
    \centering
    \includegraphics[width=0.9\linewidth]{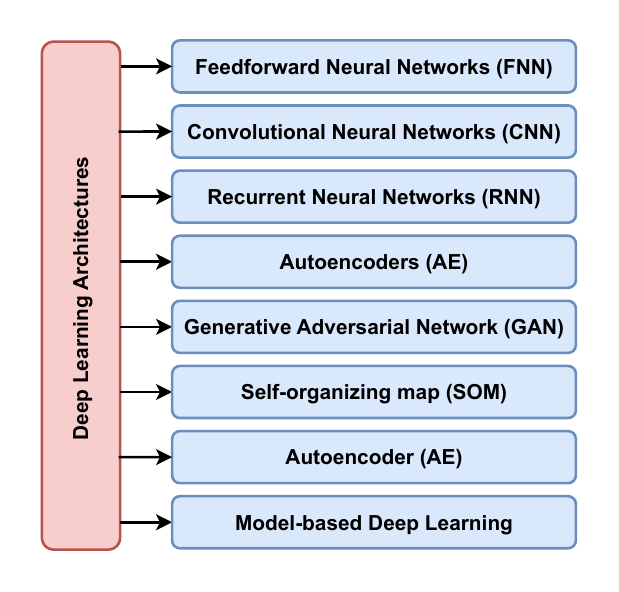}
    \caption{{DL architectures.}}
    \label{fig:deep_learning}
\end{figure}

{DL} models consist of many layers of artificial neural networks, and these layers are trained through large amounts of data during a learning procedure \cite{lecun2015deep}. More extensive neural networks can attain higher accuracy on more complex and sophisticated tasks that also require massive amounts of data for learning \cite{ahmed2023deep}. Moreover, recent advances in CPUs, GPUs, and distributed computing have enabled training more extensive neural network (NN) models with vast amounts of data\cite{goodfellow2016deep}. Fig.~\ref{fig:deep_learning} presents the widely used {DL} {architectures}, and a brief review of them is presented below\footnote{{It is worth noting that there exist more architectures that are not mentioned here since they may be used for more specific problems outside the scope of this article.}}.

The advanced algorithms and techniques proposed in the literature for communication systems can have a very high complexity or hardware complexity to be implemented in real-time systems. However, {DL} can solve this complexity issue by utilizing a data-driven \cite{dai2020deep} or model-based approaches \cite{shlezinger2023model}. {The main computational burden of DL-based  {techniques} is encountered during training, especially in data-driven approaches, since training large-scale DL models requires a large amount of data, and computational and memory resources. However, the training is usually performed offline on GPUs or distributed computing architectures \cite{mayer2020scalable}, hence, the computational complexity involved in training deep learning models does not impact their deployment. Accordingly, the trained DL models can be implemented even on low-power edge devices since the computational complexity of inference of DL models is significantly lower compared to iterative or optimization-based methods \cite{zawish2024complexity}. Moreover, task-specific fine-tuning of the DL models can be performed based on the pre-trained models via federated edge learning without introducing a significant computational burden \cite{LyuPreTraining2025}.}

The application of {DL} has recently grown in wireless communication systems due to its advantages over conventional signal processing techniques \cite{erpek2020deep}. It is expected to be a vital part of 6G wireless networks \cite{Ozpoyraz6GDeepLearning2022}. DL-based techniques can also interpret and solve high-complexity problems encountered while designing ISAC systems, such as optimum waveform design, channel estimation, interference cancellation, or receiver processing. 

\subsubsection{Feedforward Neural Networks}
Feed-forward neural networks (FNN) (also called multilayer perceptron (MLP) or fully connected deep neural networks (FCDNN)) consist of multiple layers of fully connected neurons. These models generally use linear or nonlinear activation functions to learn sophisticated patterns in the data. Each neuron in a layer is connected with activation functions to all neurons from the previous layer, and the last layer is connected to the output \cite{goodfellow2016deep}. FNN can be combined with other {DL} architectures given below to achieve a better learning performance, computational efficiency, or prediction accuracy, thus they are widely employed in communication and ISAC systems \cite{sainath2015convolutional, jiang2024isac}.

\subsubsection{Convolutional Neural Networks} Convolutional neural networks (CNN) consist of convolutional layers, which are utilized to learn specific features and patterns in the data. CNN is a powerful technique, especially for image processing. Accordingly, many modern DL-based image processing models include numerous convolutional layers in addition to fully connected ones. CNN can be utilized in ISAC systems, especially for channel estimation, since it can accurately learn and classify specific and dominant features of the channel and may outperform fully connected layers \cite{gao2023two}.  {Moreover, CNNs are widely used} in target detection, parameter estimation, and target recognition and classification in radar systems \cite{GengRadar2021}. Hence, it can also be utilized in ISAC systems to process radar images.

\subsubsection{Recurrent Neural Networks}
Recurrent neural networks (RNNs) are DL-based algorithms that strive to learn and utilize the relation between sequential data for temporal problems. RNNs take into account previous input instances in addition to the current input \cite{de2015survey, sundermeyer2015feedforward}. In RNNs, the hidden layers are composed of recurrent cells of which states are impacted by both past conditions and present information via feedback links \cite{yu2019review}. Long short-term memory (LSTM) is a class of RNNs that is developed by enhancing the standard recurrent cells with the introduction of input, output, and forget gates. Accordingly, an LSTM cell can determine which information will be discarded (forgotten) from the cell state, resulting in an improved learning performance \cite{yu2019review}. RNN and LSTM can efficiently learn datasets that have a correlation or relation within sequential data instances. For instance, channel estimation can be performed by RNN or LSTM since sequential channel instances are correlated in many wireless communication scenarios \cite{nguyen2023channel, pan2021channel, mohammed2023deep}. Moreover, LSTM is also considered for predictive beamforming design, where the beamforming matrix for the next time slot is predicted based on the historical channel data \cite{Predictive-beamforming-Vehicular-LSTM-mcf, zhang2024predictive}.

\subsubsection{Autoencoder}
An autoencoder (AE) is a specific type of artificial neural network that encodes and decodes unlabeled data efficiently via unsupervised learning. It comprises two main elements: an encoder, which transforms the input data into a compressed representation, and a decoder, which strives to reconstruct the input from this compressed representation \cite{ZhaiAutoEncoder2018}. This mechanism facilitates the autoencoder in discovering an efficient encoding method for a dataset that can be utilized to solve various problems. Autoencoders have already been considered to mitigate various conditions such as carrier frequency and phase offset or multi-path fading in wireless networks  \cite{zou2021channel, kokalj2019autoencoders}.

\subsubsection{Generative Adversarial Network}

Generative Adversarial Network (GAN) is a generative modeling approach based on {DL} architectures such as CNN or MLP. GAN models aim to recognize patterns in the input dataset to generate new samples similar to the genuine data.  GAN generally has two trainable and differentiable modules: a generator and a discriminator. The initial input of the generator is typically random noise. The training is based on game theory, such that the generator and discriminator iterate to optimize their performance. The discriminator strives to differentiate fake and real signals accurately, while the generator strives to generate fake signals increasingly resembling the real signals with each training iteration \cite{ZouGAN2024}. GAN-based techniques have been proposed for channel estimation, signal classification, data receiver, and signal denoising in wireless communication systems \cite{ZouGAN2024, shi2023data, TangDenoisingGAN2023}.

\subsubsection{Self-organizing map}
The self-organizing map (SOM) is an artificial neural network architecture based on unsupervised learning, which maps high-dimensional input data to a low-dimensional representation \cite{kohonen2013essentials}. SOM is a bioinspired ML model motivated by retina-cortex mapping with a low computational complexity. Self-organization can be used for pattern recognition, clustering, classification, data visualization{,} and mining\cite{yin2008self}. Some applications of SOM in wireless communication networks are anomaly detection in wireless networks \cite{allahdadi2021hidden}, efficient path planning for UAV-IoT systems \cite{gad2024joint}, or localization in wireless sensor networks \cite{ertin2005self}. 

\begin{table*}[ht!] 
\centering
{\caption{Deep learning architectures, their main characteristics, and example use cases in ISAC.}
\begin{tabular}{|l|l|l|}
\hline
                     \textbf{Architecture} & \textbf{Characteristics}                                                                                                                                                                        & \textbf{Example applications for ISAC }                                                                                                               \\ \hline
\textbf{MLP}          & \begin{tabular}[c]{@{}l@{}}Fully connected layers\\ General-purpose method \end{tabular}                                                                                                  & \begin{tabular}[c]{@{}l@{}}Waveform design and optimization \cite{PulkkinenMFOL2023} \\ Beamformer/precoder optimization \cite{TemizUnsupervised2025, AE2022}\\ Data modulation and demodulation \cite{wuSensingIntegratedDFTSpread2023}\end{tabular} \\ \hline
\textbf{CNN}          & \begin{tabular}[c]{@{}l@{}}Utilizes convolutional layers\\ Captures features and patterns in data\end{tabular}                                                                                   & \begin{tabular}[c]{@{}l@{}}Channel estimation\cite{LiuDLCSI2023, CSI-DRSN-mcf}\\ Target parameter estimation \cite{MateosSemi2024, suarezDeepLearningaidedRobust2023}\\ Gesture and motion recognition\cite{liuVerticalFederatedEdge2022} \end{tabular}                            \\ \hline
\textbf{LSTM-RNN}         & \begin{tabular}[c]{@{}l@{}}Remembers previous instances\\ Effective for time-series and sequential data\end{tabular}                                                   & \begin{tabular}[c]{@{}l@{}}Predictive beamforming \cite{Predictive-beamforming-Vehicular-LSTM-mcf, zhang2024predictive}\\ Data demodulation\cite{liuDeepLearningBased2022} \\
Target tracking \cite{liu2020deepmtt}\end{tabular}                                                               \\ \hline
\textbf{RL}           & \begin{tabular}[c]{@{}l@{}}Learns optimal decisions by exploration\\ Leverage feedback to dynamically optimize\end{tabular}                                                                & \begin{tabular}[c]{@{}l@{}}Resource allocation and optimization \cite{LiuResourceAl2025, li2025deep}\\ UAV trajectory planning for ISAC \cite{LinSensing2023}\end{tabular}                                \\ \hline
\textbf{Transformers} & \begin{tabular}[c]{@{}l@{}}More sophisticated architectures\\ Utilizes self-attention and relations in data\\ Resource intensive and needs larger datasets\end{tabular} & \begin{tabular}[c]{@{}l@{}}Joint data and target parameter estimation \cite{hu2024isac}\\ Predictive beamforming \cite{ZhangTransformerPred2024,ZhangTransformerRate2024}\end{tabular}                                      \\ \hline
\end{tabular}\label{tab:models}}
\end{table*}

\subsubsection{Model-based Deep Learning}
Model-based {DL} methods utilize both explicit expert knowledge of the system and a {DL} approach to yield solutions to particular problems \cite{shlezinger2023model}. Utilizing even an approximate mathematical model of the problem in combination with {DL} can significantly reduce the amount of data required compared to data-driven {DL} models. \cite{zappone2019wireless}. Furthermore,  model-based {DL} approaches also enable the DL-based solutions to be better understood and predictable compared to data-driven black-box models. Model-based {DL} integrates detailed models of the system or algorithms into the learning process. Unlike black-box {DL} approaches that learn to make predictions directly from data, model-based {DL} approaches endeavor to understand the underlying mechanisms of the model or algorithm \cite{shlezinger2023model}. One specific example of model-based {DL} is deep unfolding, which incorporates both {DL} and traditional iterative algorithms employed in signal processing or wireless communications. Deep unfolding aims to leverage the well-known structures of algorithms through learnable parameters in {DL} models \cite{mohammad2023unsupervised}. Employing model-based {DL} architectures, the computational complexity of iterative techniques can be reduced to facilitate their implementation in real-time systems.

{Table~\ref{tab:models} presents widely used DL architectures, their features, and suitable applications of them in ISAC systems. It is possible to utilize various DL architectures to develop a wide range of methods for ISAC systems. For instance, DNN and CNN architectures can be used for channel estimation. On the other hand, LSTM or transformers can be used for predictive beamforming since they capture temporal correlations and relations between the channel instances. However, the choice of the DL architecture will affect the performance and resource usage.} This section briefly reviews {DL} architectures that can be applied in wireless communication and ISAC systems.  More details regarding {DL} architectures and their applications can be found in \cite{bishop2006pattern, alpaydin2021machine, goodfellow2016deep, lecun2015deep, Ozpoyraz6GDeepLearning2022, hatcher2018survey, WangLargeScaleML2022}.  
\begin{figure*}
    \centering
\includegraphics[width=1\linewidth]{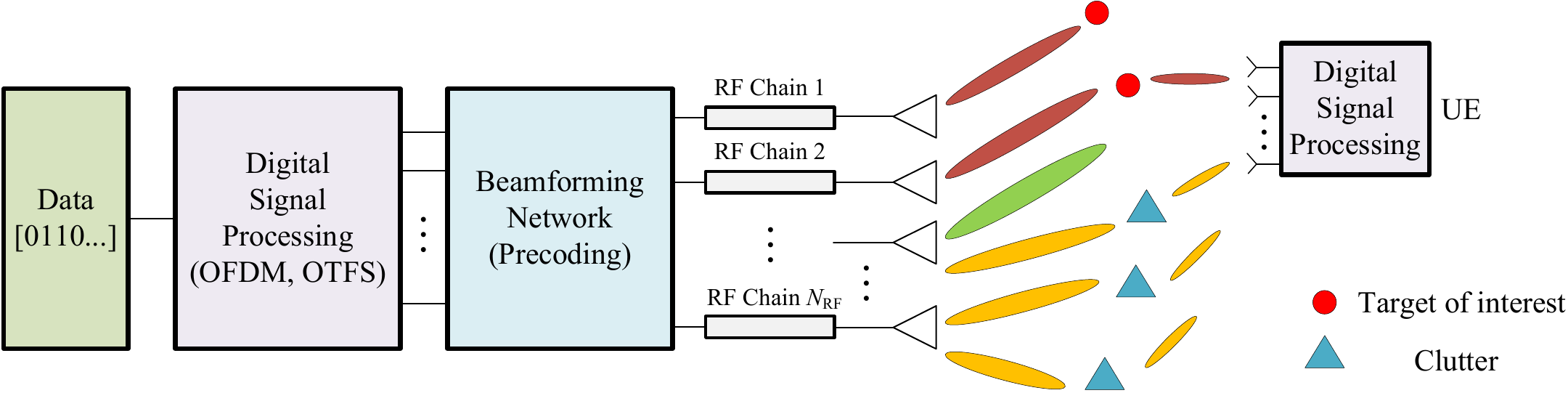}
    \caption{A communication-centric ISAC system architecture.}
    \label{fig:Comm_Centric}
\end{figure*}
\section{ISAC {Fundamentals}}
ISAC is one of the primary techniques that is expected to be at the core of 6G and beyond systems. {Various applications can employ ISAC to improve the service quality, energy efficiency, or performance while reducing the power consumption \cite{LuISACRecentChallenges2024, andersson2021joint,bayesteh2022integrated}. For instance, autonomous and intelligent vehicles can communicate with each other while sensing their environment to enable safe and reliable self-driving \cite{cheng2022integrated, beamforming2022}. This section briefly presents the fundamentals of ISAC systems with widely used performance metrics.}

\subsection{Collocated Communication and Radar Systems}

{Radar and communication systems can interfere with each other if they are closely located and operate in overlapping frequencies. The interference between communication and sensing systems significantly affects both systems and, hence, needs to be managed and coordinated to achieve the best performance of both systems. To alleviate this performance degradation, radar and communication systems need to share data such as channel state information (CSI), transmission time or frequency bands, and transmitted data to achieve coordination between them.  Coordination and cooperation between these two systems are not easy tasks since they require a massive amount of data sharing and strict policies. Consequently, a dual-function system, i.e., an ISAC system, can be developed on the same hardware platform to provide both sensing and communication functions. A wide range of techniques can be utilized to devise ISAC systems that provide both sensing and communication functions that employ the same hardware and signals. These techniques are generally considered under three major categories, as explained below.}

\subsection{Communication-centric ISAC Design}

Communication waveforms can be utilized for radar sensing, with causing a minimal impact on communication performance via various techniques. These techniques modify communication waveforms to facilitate sensing within them. Hence, the primary function of the modified waveforms is to deliver data transmission while performing sensing as a secondary function. For instance, sidelobes of the communication signals can be constructed toward possible targets, while main beams are formed for communication {UEs}. Due to being based on existing communication waveforms, these techniques are prominent candidates for 6G and beyond wireless communication networks to enable sensing within the communication network \cite{temiz2020dual, barneto2019full, ChenSensing5G2022, temiz2021dual, WanOFDMSensing2024, hawkins2024ofdm}. {As an example, a communication-centric ISAC system architecture is shown in Fig. \ref{fig:Comm_Centric}, where orthogonal frequency division multiplexing (OFDM) and orthogonal time frequency space (OTFS) waveforms can be utilized for communication, and the beamforming is performed by also considering possible target directions in addition to communication {UEs}. OFDM and OTFS are considered attractive waveforms for 6G and beyond networks \cite{sturm2011waveform, SansonOFDM2020, gaudio2020effectiveness, GaudioOTFSOFDM2022}.}

\subsection{Radar-centric ISAC Design}

\begin{figure}
    \centering
    \includegraphics[width=1\linewidth]{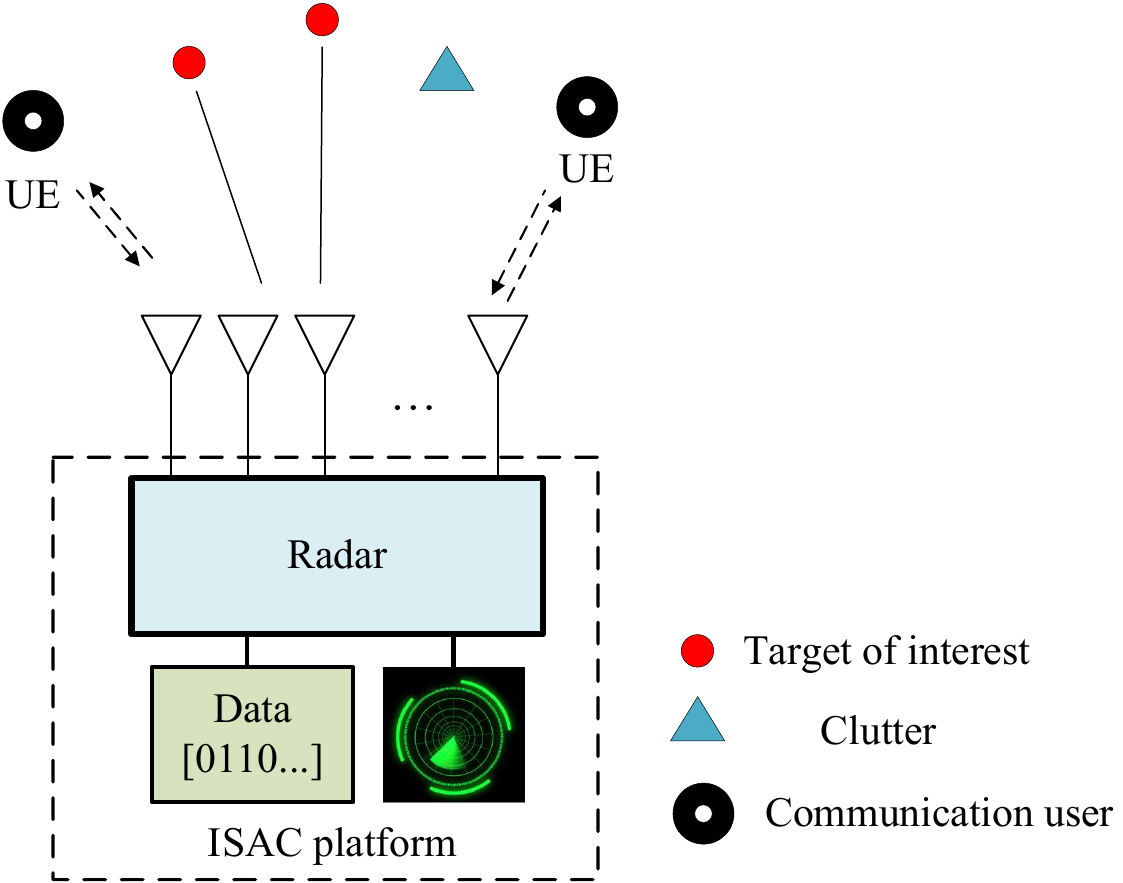}
    \caption{A radar-centric ISAC system architecture.}
    \label{fig:ISAC_Radar_Centric}
\end{figure}

It is also feasible to enable data transmission in sensing systems to communicate with {UEs}. Radar-centric ISAC techniques aim to enable communications using sensing waveforms without significantly changing the radar hardware and system complexity. In a radar-centric ISAC approach, sensing tasks are the primary concern, whereas communication is considered as a secondary function. A radar-centric ISAC system architecture is illustrated in Fig. \ref{fig:ISAC_Radar_Centric}, where the radar waveform is modified to carry data to the end {UEs}. For instance, FMCW chirps can be modified using index modulation (IM) to enable data transmission within a short-range radar \cite{TemizISACSensing2023, temiz2023radar}. { Moreover, other radar waveforms can also be used to transmit data, such as linear frequency modulated (LFM) signals, which can be modulated via minimum-shift keying (MSK) or phase-shift keying (PSK) to carry data \cite{bekar2020joint, MSK-LFM}. Enabling sensing within radar waveforms might affect the sensing functions; nevertheless, this can be alleviated by advanced signal processing methods or the implementation of {DL} techniques.}

\subsection{Dual-Function Waveform Design and Optimization}

\begin{figure}
    \centering
    \includegraphics[width=1\linewidth]{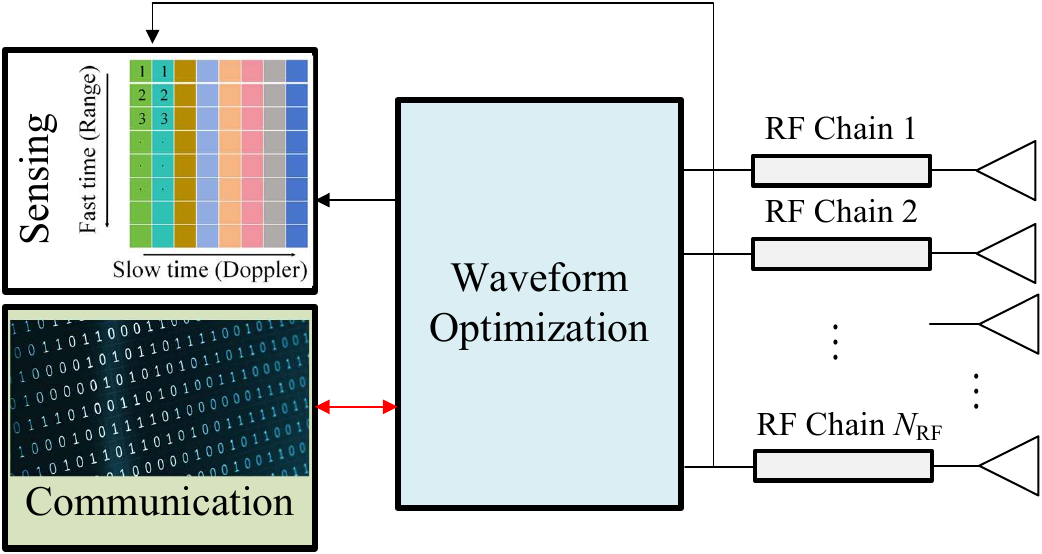}
    \caption{Design and optimization of dual-functional waveforms in an ISAC system.}
    \label{fig:ISAC_Waveform}
\end{figure}

In the communication-centric or radar-centric design of ISAC systems, one function is determined as the primary function, while the other is an auxiliary function. Consequently, those approaches can be used to improve the existing systems without significantly changing their architecture. However, the ultimate goal for ISAC systems is to have dual-functional waveforms that can provide target detection while carrying information at the same time. Such an approach can provide both functions equally without having a bias towards either of them. Moreover, the trade-off between communication capacity and sensing accuracy can be dynamically adjusted depending on instantaneous communication and sensing requirements. An approach based on a novel dual-functional ISAC waveform design is illustrated in Fig. \ref{fig:ISAC_Waveform}, where the waveform transmitted is continuously optimized by considering immediate sensing and communication requirements. Dual-functional ISAC waveforms can be designed to yield a minimized multi-user interference or constant transmit power per antenna, or total transmit power. This ISAC waveform optimization problem can then be formulated  as \cite{liu2018toward}, 
\begin{equation}
 \begin{aligned}
 &\min _{\mathbf{X}} \quad\quad\|\mathbf{H X}-\mathbf{S}\|_{F}^{2} \\
 &\text { s.t. } \quad\quad\left\|\mathbf{X}-\mathbf{X}_{0}\right\|^2_{F} \leq \eta, \\
 & \quad\quad \quad\quad\left|x_{i, j}\right|=\sqrt{\frac{P_{T}}{N}}, \forall i, j, \\
 \end{aligned}
 \label{eq:wave_sim}
\end{equation}
where $\mathbf{X} \in \mathbb{C} ^{N\times L} $, $\mathbf{S} \in \mathbb{C} ^{N\times L}$ and $\mathbf{X} _{0} \in \mathbb{C} ^{N\times L}$ denote the transmitted ISAC signal matrix, desired symbol matrix for communications and optimum radar signal matrix for sensing, respectively, where $N$ and $L$ denote the number of transmit antennas and number of symbols. Matrix $\mathbf{H} \in \mathbb{C} ^{K\times N}$ denote the CSI matrix between the $K$ {UEs} and $N$ transmit antennas. {The waveform design problem given by} \eqref{eq:wave_sim} strives to minimize the multi-user interference, {$\left\|\mathbf{HX-S}\right\|_F^2$},  while ensuring that the per-antenna power constraint is satisfied and that the difference between the transmitted matrix and the optimum radar waveform, {$\left\|\mathbf{X}-\mathbf{X}_{0}\right\|^2_{F}$}, is below a certain threshold, denoted by $\eta$, satisfying a minimum sensing performance. {Parameter $0\leq\eta$ defines a threshold for the difference between the optimum radar waveform $\mathbf{X}_0$ and the optimized ISAC waveform, such that a lower value of $\eta$ enforces a higher similarity.} {For this optimization, the optimum radar signal matrix $\mathbf{X}_0$ need to be computed in advance.}

It is also possible to devise ISAC waveforms without having any reference waveforms for sensing and communications. Such waveforms can equally provide sensing and communication performance, or a trade-off parameter between communication and sensing performances can be included in the waveform design to dynamically adjust the balance between their sensing accuracy and communication, depending on the situation or instantaneous sensing and communication requirements. For instance, design of an ISAC MIMO precoder, $\mathbf{W}$, can also be formulated as follows {\cite{TemizUnsupervised2025}},
\begin{equation}
 \begin{aligned}
 &\max _{\mathbf{W}} \quad\quad\rho\beta \sum_{k=1}^K {C_k\left(\mathbf{W}\right)} + \frac{(1-\rho)\gamma} {\sum_{t=1}^{T}{\psi_t\left(\mathbf{W}\right)}} \\
 &\text { s.t. } \quad\quad||\mathbf{W}||_F^2 \leq P_t,
 \end{aligned}
 \label{DualFunctionWaveform}
\end{equation}
where $\rho$ is the weighting parameter to adjust the trade-off between communication capacity and sensing accuracy. Moreover, ${C_k\left(\mathbf{W}\right)}$ denotes the channel capacity of the $k$th {UE}, ${\psi_t\left(\mathbf{W}\right)}$ denotes the Cramer-Rao lower bound (CRLB) of the waveform for the $t$th target, {and both these performance metrics are the functions of the precoder matrix, $\mathbf{W}$.} Parameters $\beta$ and $\gamma$ are used to normalize the sum capacity and sum CRB values, and $P_t$ denotes the total transmit power budget. 

The optimization problem (\ref{eq:wave_sim}) aims to design the ISAC waveform by shaping the transmit waveform between the optimum communication and optimum sensing waveforms. However, it does not guarantee the desired sensing and communication since it does not take into account the performance metrics. Moreover, it requires knowledge of the optimum waveforms. Hence, this problem can be solved by a supervised learning approach. On the other hand, the optimization problem (\ref{DualFunctionWaveform})  aims to design a novel waveform that can satisfy both sensing and communication performance metrics. Moreover, the latter method does not require knowledge of the optimum waveforms; hence, it can be optimized via an unsupervised learning approach.  


\subsection{Performance Metrics for ISAC}

The sensing capability of a radar and the sum rate of a communication system are bounded by waveform parameters, which are functions of waveform characteristics such as bandwidth, signal-to-interference-plus-noise ratio (SINR), and other signal features. Radar range and velocity estimation accuracies of a radar signal are predominantly determined by its ambiguity function, which is a widely used performance metric for sensing. Let $u\left(t\right)$ define a signal that is used for sensing; its ambiguity function (AF) is then given by \cite{Abatzoglou1998} 
\begin{equation}
\Omega\left(\tau,\omega\right)=\int_{-\infty}^{+\infty}u\left(t\right)u^{*}\left(t-\tau\right)e^{-j2\pi\omega t}dt,
\end{equation}
where $\tau$ and $\omega$ denote delay and Doppler frequency caused by the sensing of targets, respectively. 

{Another widely used sensing performance metric is CRLB, which is calculated for target parameter estimation, e.g., range, velocity, angle, localization, in radar systems \cite{liu2021cramer, Greco2011}. The generalized equation for CRLB for the estimation of a deterministic parameter is given by \cite{LiuSurveyisac2022}
\begin{equation}
\text{CRLB}_{\boldsymbol{\theta}}=\mathbf{J}^{-1}(\boldsymbol{\theta}),
\end{equation}
where $\mathbf{J}(\boldsymbol{\theta})$ denotes the Fisher information matrix (FIM) and its ($i$,$j$)-th element is given by 
\begin{equation}
[\mathbf{J}(\boldsymbol{\theta})]_{i,j}=\mathbb{E}\left[\frac{\partial \ln p(\boldsymbol{y}; \boldsymbol{\theta})}{\partial \theta_{i}} \frac{\partial \ln p(\boldsymbol{y}; \boldsymbol{\theta})}{\partial \theta_{j}}\right],  
\end{equation}
where  $p(\boldsymbol{y}; \boldsymbol{\theta})$ is the likelihood function associated with estimating the parameter vector $\boldsymbol{\theta}$ from the measurements $\boldsymbol{y}$. Moreover, the probability of detection and false alarm rate are also widely used as sensing performance metrics.}

\begin{figure*}[t]
    \centering
    \includegraphics[width=0.7\linewidth]{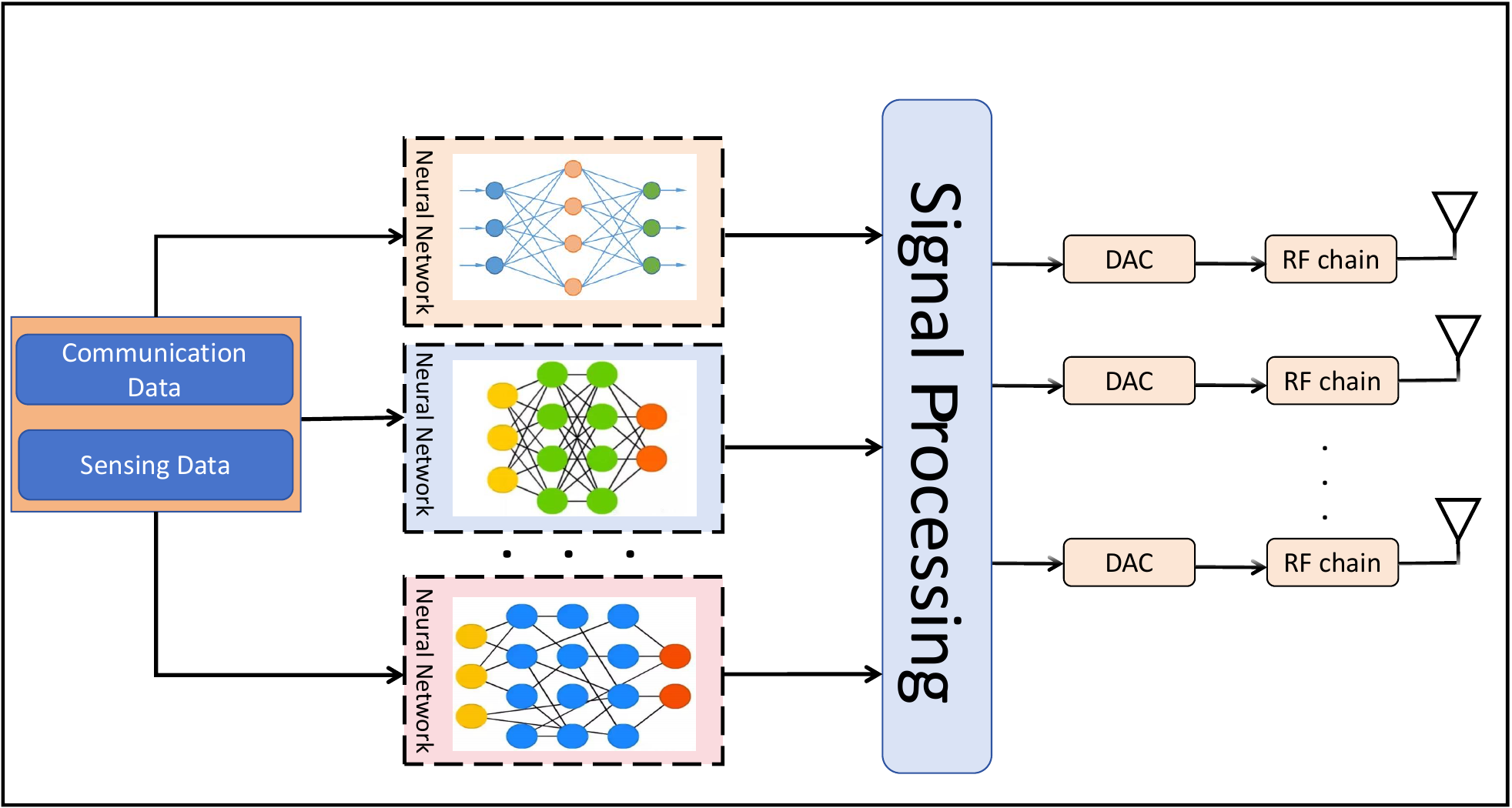}
    \caption{DL-based ISAC transmitter architecture.}
    \label{fig:ISAC_transmitter}
\end{figure*}

On the other hand, performance metrics widely used for communication systems are bit error rate (BER), block error rate (BLER), symbol error rate (SER), sum rate, and energy efficiency. For instance, the sum rate of a network consisting of $K$ communication {UEs} is given by 
\begin{equation}
    R_K = \sum_{k=1}^K \log_2\left(1 + \frac{p_k}{i_k + n_k} \right),
\end{equation}
where $p_k$, $i_k$, and $n_k$ denote the received useful signal power, interfering signal power, and noise power at the $k$th UE. The ISAC systems aim to maximize both communication and sensing performance metrics regarding a desired trade-off between them within the system limitations.

\section{DL-based Techniques for ISAC Transmitters}

\begin{table*}[]
    \centering
    \caption{DL-based techniques for ISAC transmitters.}
    \footnotesize
    \begin{tabular}{|l<{\centering}|m{2cm}<{\centering}|m{2cm}<{\centering}|m{2cm}<{\centering}|m{2.5cm}<{\centering}|m{3cm}<{\centering}|m{3cm}<{\centering}|}
\hline  
            \textbf{Ref.} & \textbf{Learning Strategy} & \textbf{Algorithm} & \textbf{Inputs} & \textbf{Outputs} & \textbf{Pros} & \textbf{Notes} \\
            \hline
            \cite{PulkinenISAC2024} & Model-based Online  RL & MLP with ReLU & CSI, power budget & PA, SCS, optimized ISAC waveform & \begin{itemize}[leftmargin=4pt]
                \item Effective under interference
                \item Data-efficient
                \item Model-based approach
            \end{itemize} &  
            \begin{itemize}[leftmargin=4pt]
                \item Requires specific assumptions
                \item Data-driven method may perform better with a large amount of data
            \end{itemize}   \\
            \hline

            \cite{PulkkinenMFOL2023} & Model-free  Online RL & MLP with ReLU & CSI, power budget & PA, SCS, optimized ISAC waveform  & 
            \begin{itemize}[leftmargin=4pt]
                \item Outperforms model-based method with a large number of training samples.
            \end{itemize} &  
            \begin{itemize}[leftmargin=4pt]
                \item Data-driven method
                \item Slower convergence
            \end{itemize}   \\
            \hline
            \cite{zheng2024endtoend} & Unsupervised Learning & MLP with ReLU \& LSTM & Comm data, CSI, Target angles & SLP ISAC beamformer, comm data, target detection &
             
            \begin{itemize}[leftmargin=4pt]
                \item End-to-End Learning for the entire ISAC system
                \item Joint waveform design
                \item Satisfactory performance
                \item Low-complexity SLP
            \end{itemize} &
            \begin{itemize}[leftmargin=4pt]
                \item MLP for transmitter, receiver, and target detector
                \item LSTM for target angle estimation
            \end{itemize}  \\
            \hline
            \cite{RL2023} & RL  & FCDNN & Waveform parameters & OCDM waveform & 
            \begin{itemize}[leftmargin=4pt]
                \item Enhanced target detection
                \item Enhanced symbol decoding
            \end{itemize} & 
            \begin{itemize}[leftmargin=4pt]
                \item Joint waveform design, target detector, and comm decoder
                \item Data-driven method
                \item Supervised learning for the receiver
            \end{itemize} \\
            \hline
            \cite{AE2022} & Supervised End-to-end Learning & Autoencoders (FCDNN) & Communication data, possible locations of targets and CSI & ISAC beamformer & 
            \begin{itemize}[leftmargin=4pt]
                \item Robust to hardware impairments
                \item Similar performance to the benchmark
            \end{itemize}
            
            & \begin{itemize}[leftmargin=4pt]
                \item Data-hungry
                \item End-to-end training
                \item Data-driven method
                \item Long training duration
            \end{itemize}\\
            \hline
               \cite{XuRecon2024} & RL &  DRL (FCDNN) & 
               Estimated ranges, angles, and velocities of targets 
               & Beamforming weights and antenna selection
               & 
            \begin{itemize}[leftmargin=4pt]
                \item Achieves the desired beam pattern
                \item Outperforms  the benchmark
            \end{itemize}
            
            & \begin{itemize}[leftmargin=4pt]
                \item DRL-based method
                \item ISAC for automotive
                \item High computational complexity
            \end{itemize}\\
            \hline
            \cite{liu2023distributed} & Unsupervised \& Transfer Learning & 
                 \centering FCDNN
               
               & \centering CSI
               
               & Beamforming matrix &
            \begin{itemize}[leftmargin=4pt]
                \item Transfer learning is used to reduce training time
                \item Outperforms supervised learning 
            \end{itemize}
            
            & \begin{itemize}[leftmargin=4pt]
                \item Transfer learning
                \item Distributed interference management
            \end{itemize}  \\
            
            \hline
\cite{TemizUnsupervised2025} & {Unsupervised Learning} &  \centering FCDNN
               & \centering CSI, target angles, and trade-off parameter
               & Precoding matrix &
            \begin{itemize}[leftmargin=4pt]
                \item Adjustable trade-off with multiple DNNs
                \item Low complexity via weight quantization and pruning
                \item Outperforms WMMSE
            \end{itemize}
            & \begin{itemize}[leftmargin=4pt]
                \item Low memory usage
                \item Trade-off design
            \end{itemize}  \\
            \hline
        \end{tabular}
        \label{tab:ML-transmitter}%
    \end{table*}

Signal design is essential to developing ISAC systems since the signals need to be carefully designed to deliver both sensing and communication functions. DL-based techniques can significantly reduce the computational complexity of ISAC signal design. The ISAC transmitter mainly deals with modulating the communication data, generating and optimizing the ISAC waveform, and designing beamforming or precoding in MIMO systems. {DL} methods can be primarily utilized for digital signal processing in ISAC transmitters, as shown in Fig.~\ref{fig:ISAC_transmitter} to replace the computationally intensive iterative or optimization-based signal processing techniques. For instance, waveform design and optimization for sensing and communication, as well as precoder design and data modulation, can be performed {by DL-based techniques since they can efficiently solve complex and computationally intensive optimization problems \cite{DahroujML21, LiLearning2023}.} 


Table~\ref{tab:ML-transmitter} presents an overview of {DL} methods that have been developed for transmitters. This table shows that DL-based techniques are mainly used for ISAC waveform design and optimization, predictive beamforming, symbol-level precoding, power, and subcarrier allocation. These methods are especially proposed for communication-centric or dual-function ISAC  systems.


\begin{figure}[!h]
    \centering
\includegraphics[width=1\linewidth]{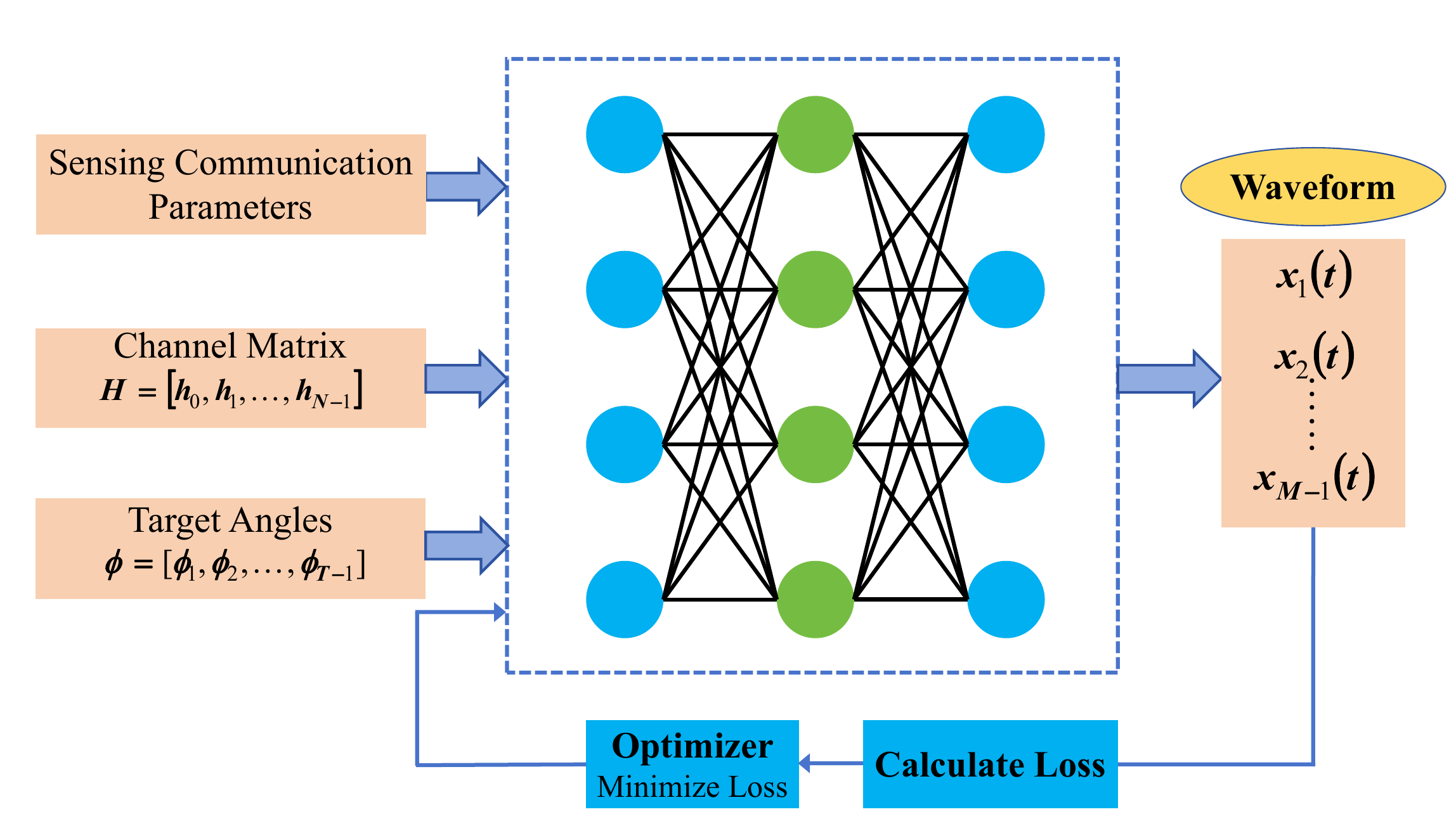}
    \caption{ISAC waveforms design via {DL}.}
    \label{fig:waveform_design}
\end{figure}

\subsection{Waveform Design and Optimization}
Fig.~\ref{fig:waveform_design} illustrates an approach to design and optimize ISAC waveforms based on {DL} techniques, where the training can be performed via supervised learning, unsupervised learning, or {RL} strategies. Moreover, model-free or model-based {DL} architecture may be employed in the design. Waveform design chiefly requires the optimization of waveform parameters or resource allocation to achieve the best performance for sensing and communications. 

Model-based and model-free online {RL} are considered to optimize power and subcarrier allocation in \cite{PulkinenISAC2024, PulkkinenMFOL2023}. The model-based method exploits the expert knowledge of the ISAC system. Hence, it requires fewer training samples and provides a highly effective solution under various interference scenarios for ISAC waveform design and optimization \cite{PulkinenISAC2024}. On the other hand, the model-free method requires many more training samples as it is a data-driven method \cite{PulkkinenMFOL2023}. It was shown that the model-based method converges much faster than the model-free method; however, the model-free method may outperform the model-based method when a large number of training samples are available. In these studies \cite{PulkinenISAC2024, PulkkinenMFOL2023}, {RL} with a single hidden layer FNN is used. Moreover, these methods eliminate the need to solve non-convex optimization problems while achieving near-optimum waveform optimization. Another study proposed a similar approach ({RL} with FCDNN architecture) to design an orthogonal
chirp-division multiplexing (OCDM) ISAC waveform in a data-driven manner \cite{RL2023}. This study modeled the entire system, including the transmitter, communication receiver, and radar receiver, as a neural network, hence utilizing an end-to-end approach to train the {RL} model.

\subsection{{Beamformer and Precoder Design for ISAC}}
Designing an optimum beamformer or precoder can be challenging due to requiring non-convex optimization or iterative methods, even for communication systems. For instance, DL-based techniques have already been studied for designing the weighted minimum mean-square error (WMMSE) beamformer for only communication systems to reduce its computational complexity \cite{chowdhury2021unfolding,shi2023robust}. Beamformer design for ISAC systems is much more complex due to involving a larger parameter space, including the parameters related to sensing (e.g., CRLB, probability of detection, radar SNR). Accordingly, DL-based  {techniques} may be efficiently utilized to reduce this complexity. Designing a symbol-level precoder (SLP) for ISAC using MLP and LSTM is proposed in \cite{zheng2024endtoend}, where an end-to-end data-driven unsupervised learning approach is implemented. This study utilizes an MLP network for SLP design, while it utilizes an LSTM model for target angle estimation.  Their simulation results revealed that the method proposed in \cite{zheng2024endtoend} achieves a satisfactory performance while significantly reducing the computational complexity caused by the SLP design for ISAC. 

Another study in \cite{AE2022} proposes an autoencoder-based method with supervised learning for ISAC beamforming design, which presented a performance similar to the computationally expensive iterative method, with a robust performance under hardware impairments. However, due to being a data-driven supervised learning method, this method requires a large number of training samples (e.g., $2\times10^7$).Moreover,  beamforming for automotive ISAC systems is explored in \cite{XuRecon2024}, where a DRL approach is adopted to select the antennas and beamforming weights. Their method was shown to outperform the relaxed optimization problem solution. However, the computational complexity of DRL for real-time operations on automotive radar hardware remains a challenge, requiring further studies. As can be seen from the studies reviewed above, supervised learning, unsupervised learning, and {RL} approaches can be used for ISAC waveform and beamformer design. Transfer learning is also utilized for ISAC beamforming design, where a pre-trained model is utilized for distributed interference management and beamforming, reducing the training time \cite{liu2023distributed}. Moreover, this enables each BS to utilize the local CSI for training in a network. {A recent study has proposed an ISAC precoding method based on unsupervised learning, which optimizes the precoding matrix to maximize the sum rate and minimize CRLB on target angle estimation for a given trade-off between the communication and sensing \cite{TemizUnsupervised2025}. It utilizes multiple FCDNNs to adjust the trade-off instantaneously. Hence, it achieves a higher sum rate than the WMMSE precoder when the trade-off parameter is adjusted to maximize the communication performance. Furthermore, it also reduces the complexity and memory usage of the DNN via weight pruning and quantization.}

\begin{table*}[]
    \centering
    \caption{DL-based ISAC predictive beamforming techniques.}
    \footnotesize
    \begin{tabular}{|l<{\centering}|m{1.5cm}<{\centering}|m{2cm}<{\centering}|m{2cm}<{\centering}|m{2.5cm}<{\centering}|m{3cm}<{\centering}|m{3cm}<{\centering}|}
\hline  
            \textbf{Ref.} & \textbf{Learning Strategy} & \textbf{Algorithm} & \textbf{Inputs} & \textbf{Outputs} & \textbf{Pros} & \textbf{Notes} \\

            \hline
                \cite{beamforming2022} & Unsupervised Learning  & CNN \& LSTM & Historically estimated CSI & Predictive beamforming matrix &
            \begin{itemize}[leftmargin=4pt]
                \item Predicts the  beamforming matrix of the next time slot
                \item Achieves a satisfactory sum rate 
                \item Does not require explicit channel tracking
            \end{itemize} & 
            \begin{itemize}[leftmargin=4pt]
                \item Data-driven method
                \item Applicable to vehicular networks 
            \end{itemize}\\
        \hline
            
            \cite{LiuCLSTM2022}   \cite{LiuPreBeam2022} & Supervised Learning  & CNN \& LSTM & Historical estimated angles &  Estimated beamforming
            angle &
            \begin{itemize}[leftmargin=4pt]
                \item Comparable performance to perfect beamforming with real-time angles
                \item Spatial and temporal features of the channel  
                \item Vehicular communication scenarios
            \end{itemize} & 
            \begin{itemize}[leftmargin=4pt]
                \item CNN and LSTM are utilized to exploit both temporal and spatial features.  
            \end{itemize}\\
        \hline

        \cite{ZhangTransformerPred2024}  & Unsupervised Learning  & Transformer (Encoder: CNN  Decoder: CNN) & Previous beamforming matrix and previous vector of signal echoes &  Beamforming matrix for next time slot &
            \begin{itemize}[leftmargin=4pt]
                \item Higher sum rate than state-of-the-art beamforming methods
                \item Reduced signaling overhead  
                \item  Transformer-based beamforming scheme
            \end{itemize} & 
            \begin{itemize}[leftmargin=4pt]
                \item  A trade-off observed between communication and sensing performance.
            \end{itemize}\\
        \hline

        \cite{wang2023intelligent}  & Supervised Learning  &  CNN \& LSTM & Vector of echo signals &  Estimated angle &
            \begin{itemize}[leftmargin=4pt]
                \item  Angular tracking based on the echo signals
                \item Higher sensing performance compared to other CNN and LSTM models  
            \end{itemize} & 
            \begin{itemize}[leftmargin=4pt]
                \item Target angle tracking and beam prediction are performed.
            \end{itemize}\\
        \hline

        \cite{ZhangIntegratedProactive2024}  & Supervised Learning  & Residual neural network (ResNet) \& FCDNN \& YOLOv5 & CSI and images &  Weighted Multi-model Feature &
            \begin{itemize}[leftmargin=4pt]
                \item Solid angle prediction accuracy, achievable rate, and outage performance
                \item Strong robustness against vehicle drifting and environmental interference
            \end{itemize} & 
            \begin{itemize}[leftmargin=4pt]
                \item Multi-modal environment information is fused and used.
            \end{itemize}\\
        \hline

        \cite{zhang2024predictive}  & Unsupervised Learning  & CNN \& LSTM & Historical CSI &  Beamforming matrix &
            \begin{itemize}[leftmargin=4pt]
                \item Satisfactory performance in
terms of both communication and sensing tasks
                \item Adjustable trade-off between sensing and communication performance.
            \end{itemize} & 
            \begin{itemize}[leftmargin=4pt]
                \item  ISAC beamforming for vehicles exhibiting complex behaviors. 
            \end{itemize}\\
        \hline

        \cite{Mubeamforming2021}  & Supervised Learning  & FCDNN & Received signal samples &  Beamforming angle &
            \begin{itemize}[leftmargin=4pt]
                \item Better communication performance than the factor graph approach
                \item Low latency and high accuracy beam tracking 
            \end{itemize} & 
            \begin{itemize}[leftmargin=4pt]
                \item Maintains reliable communications in high-mobility vehicular networks
            \end{itemize}\\
        \hline

        \end{tabular}
        \label{tab:DL-predictive}%
    \end{table*}

\subsection{Predictive Beamforming}
Predictive beamforming is a recent approach that leverages the knowledge of the historical channel data to reduce the dependency on instantaneous CSI and signaling overheads, improve the angle estimation accuracy, or target tracking in millimeter-wave systems \cite{beamforming-Bayesian-mcf}. Predictive beamforming aims to estimate the beamforming vector of the next time slots based on historical channel data or sensing information. Table~\ref{tab:DL-predictive} compares the DL-based predictive beamforming studies, where supervised learning or unsupervised learning is used to train the models.  

A combination of CNN and LSTM algorithms is generally used for predictive beamforming \cite{beamforming2022, LiuCLSTM2022, LiuPreBeam2022, wang2023intelligent, zhang2024predictive}. While CNN learns the patterns in the historically estimated CSI, the LSTM network exploits the correlation between the CSI of sequential time slots to predict the beamforming matrix in the next time slot. This method is proposed especially for vehicular networks, and it is a data-driven approach trained via unsupervised learning. {For instance, Fig.~\ref{fig:predictive_beamforming} shows that the HCL-Net predictive beamforming method, which employs both LSTM and CNN, outperforms LSTM-based, FCDNN-based, and CNN-based methods in terms of sum rate and can reach the upper bound since it can exploit both spatial and temporal features of the communication channels for beamforming \cite{beamforming2022}.On the other hand, residual neural network (ResNet), and transformer-based methods are also proposed for predictive beamforming \cite{ZhangTransformerPred2024, ZhangIntegratedProactive2024, Mubeamforming2021}.In addition to historical channel data, some studies also utilized camera images of the scenario or reflect signal echoes to improve the beamforming performance \cite{ZhangTransformerPred2024, ZhangIntegratedProactive2024, wang2023intelligent}.

\begin{figure}
    \centering
    \includegraphics[width=1\linewidth]{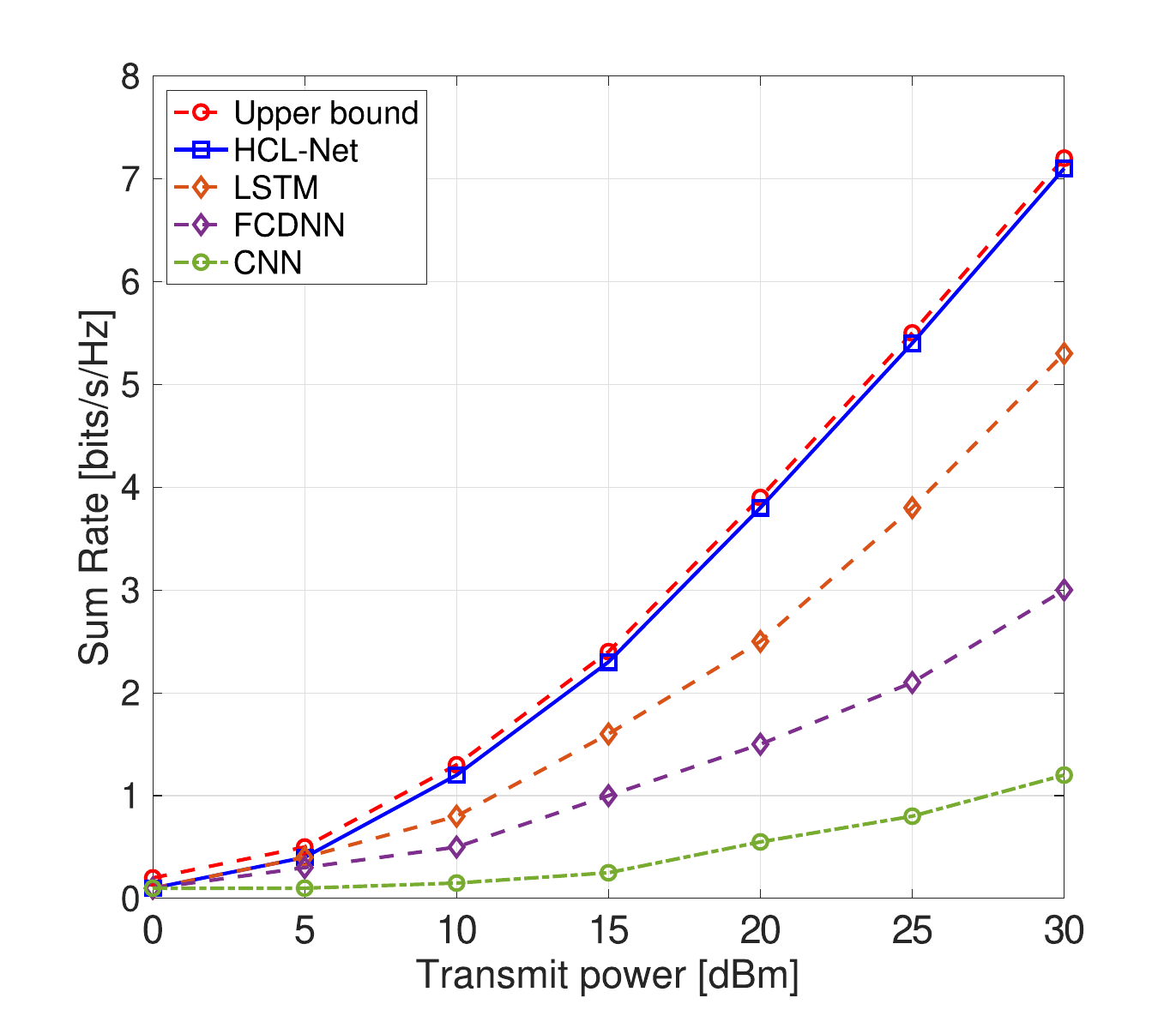}
    \caption{{Comparison of DL-based predictive beamforming methods \cite{beamforming2022}.}
    }\label{fig:predictive_beamforming}
\end{figure}

Vehicular networks can particularly benefit from DL-based predictive beamforming since obtaining the instantaneous channel data is challenging in such scenarios due to the high velocity of vehicles. Predictive beamforming methods can perform satisfactorily even in complex vehicular network scenarios since they may fuse sensing data, camera images, and historical channel data to achieve a high-accuracy beamforming performance \cite{ZhangIntegratedProactive2024, ZhangTransformerPred2024}. Predictive beamforming reduces the communication overheads and eliminates the need for instantaneous CSI. Thus, it is one of the major application areas of DL-based techniques in ISAC transmitters.

\begin{table*}
\centering
\caption{List of DL-based CSI estimation techniques for ISAC systems.}
\footnotesize
\begin{tabular}{|l<{\centering}|m{1.5cm}<{\centering}|m{2cm}<{\centering}|m{2cm}<{\centering}|m{2.5cm}<{\centering}|m{3cm}<{\centering}|m{3cm}<{\centering}|}
\hline  
\textbf{Ref.} & \textbf{Learning Strategy} & \textbf{Algorithm} & \textbf{Inputs} & \textbf{Outputs} & \textbf{Pros} & \textbf{Notes}  \\            
\hline 
\cite{LiuCSI2022}\cite{LiuDLCSI2023} &Supervised Learning & FCDNN \& CNN &The received communication and sensing signals, LS estimation of channels &   The communication and sensing channels &
\begin{itemize}[leftmargin=4pt]
    \item  Estimates  direct SAC channel, reflected communication channel, and reflected sensing channel
    \item Better NMSE performance compared to the LS estimator 
    \item Reasonable complexity
\end{itemize}
& 
\begin{itemize}[leftmargin=4pt]
\item There-stage DNN-based estimator for direct SAC channel, IRS reflected communication channel, and IRS reflected sensing channel
\item Decent generalization capacity for various SNRs
\item Data-driven method
\end{itemize}\\
\hline  
\cite{CSI-DRSN-mcf}& Supervised Learning & FCDNN \& CNN & Received pilot signals & The communication and sensing channels & 
\begin{itemize}[leftmargin=4pt]
    \item  Outperforms traditional estimators for OTFS communication.
    \item Provides a better sensing performance in terms of false alarm probability.
\end{itemize}
&\begin{itemize}[leftmargin=4pt]
    \item Utilizes denoising  and feature elimination modules
    \item Data-driven method
    \item Requires a large number of training samples
\end{itemize}\\
\hline  
\cite{LiuEL_CSIA2023}&Supervised Learning& ELM & The 
received Sensing and Communication signals & The communication and sensing channels &
\begin{itemize}[leftmargin=4pt]
    \item Reduced computational complexity 
    \item Extremely training compared to CNN-based estimators
    \item Comparable estimation performance to FCDNN-based estimators
\end{itemize}
& \begin{itemize}[leftmargin=4pt]
\item Practical scenarios and self-interference need to be considered.
\item Data-driven method.
\end{itemize} \\
\hline

\end{tabular}
\label{tab:DL-channel-estimation}
\end{table*}

\section{DL-based Channel Estimation}

\begin{figure}[]
    \centering
    \includegraphics[width=1\linewidth]{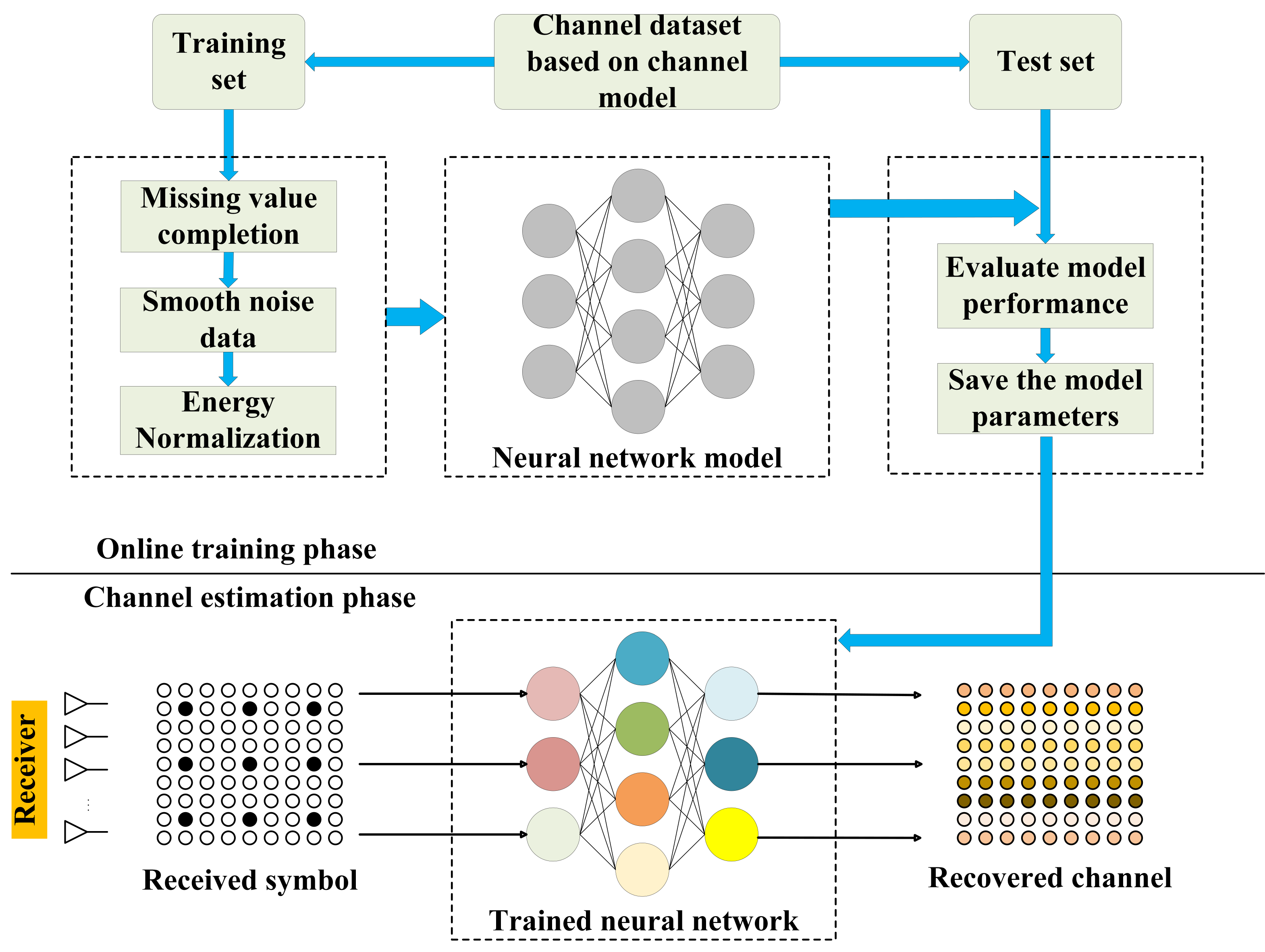}
    \caption{An example of DL-based channel estimator.}
    \label{fig:CSI-estimation}
\end{figure}

Channel estimation and equalization are essential parts of communication systems since the transmitted signals are distorted and reshaped by the channel. Accordingly, the CSI must be frequently estimated, and its impact on the received signals must be removed by channel equalization. CSI estimation is typically performed by transmitting pilot signals within the communication data frame. The receiver estimates the channel by utilizing these pilot signals and performs channel equalization to remove its impact on the received signal. In ISAC systems, the information about the propagation environment obtained by sensing can also be used for channel estimation, or the pilot symbols can be utilized for sensing.

\subsection{Conventional Channel Estimation}

Widely used channel estimation techniques are linear methods such as least squares (LS) or linear minimum mean-square error (LMMSE) estimation \cite{liu2014channel}. However, such linear methods may not be sufficient for ISAC systems because they perform both sensing and communication at the same time, and the interference in received pilot signals due to the sensing can be more complex.

To exemplify, the LS channel estimation strives to solve the following optimization problem to estimate the channel {in massive MIMO communication systems \cite{neumann2015channel} as,
\begin{equation}
   \hat{\mathbf{H}}_{LS} =  \mathop{\arg\min}_{\mathbf{H}_c} \|\mathbf{Y}_p-\mathbf{H}_c\mathbf{X}_p\|^2 = \mathbf{Y}_p \mathbf{X}_p^\dagger,
\end{equation}
where $\mathbf{H}_c\in \mathbb{C}^{M\times K}$ denotes the communication CSI matrix for $M$ antennas and $K$ {UEs}, $\mathbf{X}_p\in \mathbb{C}^{K\times L}$, denotes the pilot symbols matrix consisting of $L$ training symbols transmitted by $K$ {UEs}. and $\mathbf{Y}_p\in \mathbb{C}^{M\times L}$ is the  signal matrix received by $M$ antennas during the channel estimation. Moreover, $\mathbf{X}_p^\dagger$ denotes the pseudo-inverse of the pilot data matrix.} Moreover, iterative methods can attain a higher channel estimation accuracy compared to linear estimators \cite{nicoli2007soft}. 

\subsection{DL-based Channel Estimation for Communications}

Fig.~\ref{fig:CSI-estimation} illustrates a generic system model for DL-based channel estimation, where the trained {DL} network estimates the channel using the pilot symbols transmitted in the data. The training of the {DL} network can be performed via supervised learning. Supervised DL-based channel estimation can be performed by utilizing the ground-truth CSI. Let $\Theta=\{w,b\}$ represent all parameters of the DL network, where $w$ and $b$ are the weights and parameters of the neuron, respectively. During the training stage, the model can use MSE as the loss function $\mathcal{L}(\Theta)$, which is expressed as:
\begin{equation}
\mathcal{L}(\Theta)=\frac{1}{N_{train}}\sum_{n=1}^{N_{train}}\|\hat{\mathbf{H}}_{DL}^{(n)}-\mathbf{H}^{(n)}\|^{2},
\end{equation}
where $N_{train}$ is the number of CSI instances in the training data set, and $\mathbf{H}$ and $\hat{\mathbf{H}}_{DL}$ denote the ground truth CSI (or estimated CSI via another method) and DL estimated CSI, respectively.

\begin{table}
\centering
{\caption{Comparison of DL-Based channel estimation in communication and ISAC systems.}
\small
\begin{tabular}{|p{1.5cm}|p{2.5cm}|p{3.5cm}|}
\hline
\textbf{Criterion} & \textbf{Communication} & \textbf{ISAC} \\
\hline
Output & CSI & CSI, target parameters \\
\hline
Training data & Pilots, CSI & Pilots, CSI, range-Doppler maps, range-angle maps, target parameters \\
\hline Metrics & NMSE, BER, SER & NMSE, BER, SER, parameter estimation accuracy (e.g., range/velocity/angle RMSE) \\
\hline
\end{tabular}
\label{tab:dl_ce_comparison}}
\end{table}

DL-based channel estimators for OFDM systems are proposed in \cite{mei2021performance, jebur2021efficient}. DNN is employed via supervised learning in \cite{mei2021performance}, where it is shown that the DNN-based channel estimator can slightly outperform the LMMSE estimator when the number of training samples is high; however, when the number of training samples is limited, it is performance is degraded. A different approach is employed in \cite{jebur2021efficient}, where the training of the DNN is performed during the training frame before data transmission. This method has been shown to outperform the LMMSE estimator substantially, and its computational complexity is much lower than that of the LMMSE estimator \cite{jebur2021efficient}. 

The channel estimation is much more challenging in the applications of massive MIMO,  IRS, ISAC, and OTFS waveforms. Accordingly, DL-based channel estimators are proposed for massive MIMO systems \cite{balevi2020massive, ma2020data}, mm-wave MIMO communications \cite{he2018deep}, IRS-powered communication \cite{yu2022deep, gao2023two,zhang2023self}, OTFS communication systems \cite{payami2024deep}. Furthermore, DL-based  {techniques} are also proposed for channel tracking in mm-wave vehicular communication and UAV communication networks \cite{moon2020deep, yu2022deep}.

\subsection{DL-based Channel Estimation for ISAC}
{DL-based channel estimation for communication systems only aims to estimate the channel between the base station and {UEs}, while ISAC systems can estimate both the communication channel and sensing channel that consists of information about target parameters such as velocity, range, and angle \cite{LiuCSI2022,CSI-DRSN-mcf,LiuEL_CSIA2023}. Accordingly, Table~\ref{tab:dl_ce_comparison} compares DL-based channel estimation in communication systems and ISAC systems. Moreover, the information obtained about the environment can be utilized to improve the communication channel estimation in ISAC systems \cite{QingSensingAided2024}.}  Thus, ISAC systems require a unique channel estimation approach since their signals are designed for both communications and sensing. Accordingly, DL-based channel estimators are also proposed specifically for ISAC systems. Table~\ref{tab:DL-channel-estimation} {compares} DL-based channel estimation studies for ISAC systems. Liu et al. proposed a DL-based channel estimator for IRS-assisted ISAC systems, where FCDNN and CNN are utilized to estimate the sensing channel and communication channel, respectively \cite{LiuCSI2022, LiuDLCSI2023}. They trained the proposed DNN networks via supervised learning such that pilot signals are inputs and ground-truth channels are the output. This study utilized three DNNs to estimate direct sensing and communication (SAC) channels, IRS-reflected communication, and IRS-reflected sensing channels. Their proposed method outperforms the LS estimator by up to 15dB and 5dB to estimate sensing and communication channels, respectively. 

Another DL-based CSI estimator is proposed for OTFS-assisted ISAC systems \cite{CSI-DRSN-mcf}, where CNN and FCDNN layers are employed and trained via supervised learning for communication and sensing channel estimation. This study also employs CNN-based denoising and feature elimination blocks to achieve a better channel estimation performance. As a result, it outperforms sparse Bayesian learning (SBL) and LMMSE estimators regarding sensing and communication performance. 

Extreme learning machine (ELM) is also considered for channel estimation in IRS-assisted multi-user ISAC systems \cite{LiuEL_CSIA2023}. ELM network is an FNN with a single hidden layer that updates its parameters via an inverse operation instead of gradient-based backpropagation, hence achieving swift training and learning. ELM network is trained via supervised learning using the {channel dataset. The communication channel estimation performance of the ELM-based estimator is similar to FCDNN-based estimator, and it outperforms the CNN-based and LS estimators, as shown in Fig.~\ref{fig:nmse_ce}. On the other hand, its training time is 20\% and 2.5\% of the FCDNN-based and CNN-based ones, receptively \cite{LiuEL_CSIA2023}.} 

\begin{figure}
    \centering
\includegraphics[width=1\linewidth]{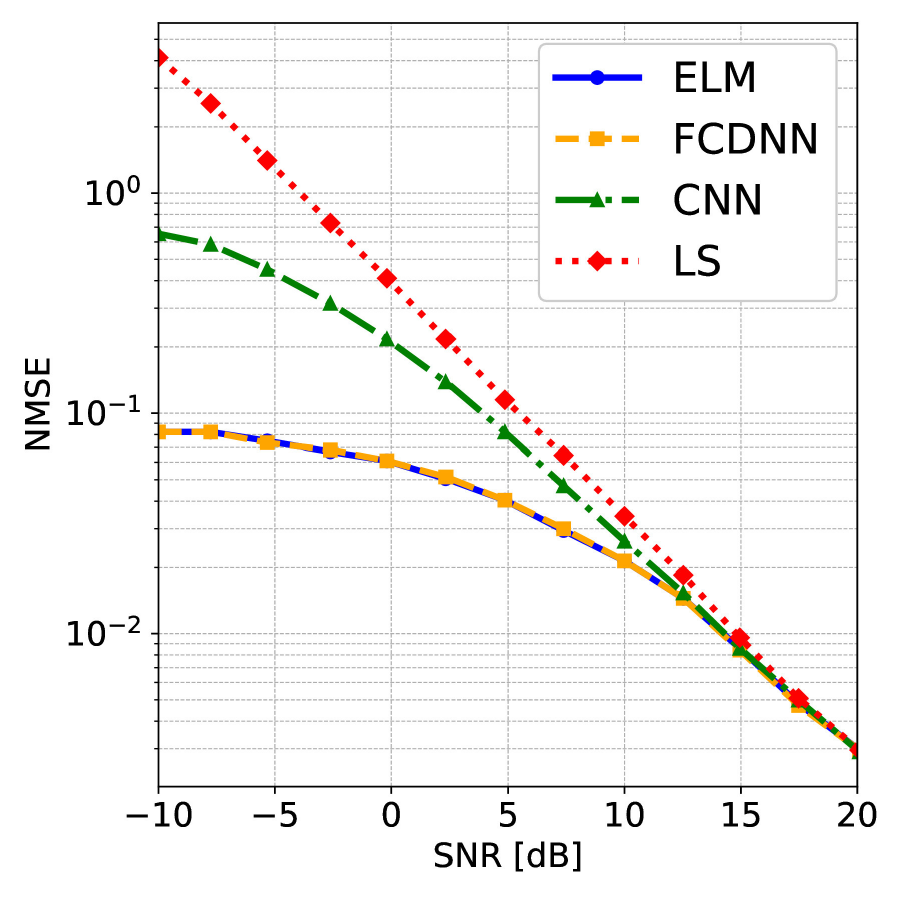}
    \caption{{Comparison of DL-based channel estimation techniques \cite{LiuEL_CSIA2023}.}}
    \label{fig:nmse_ce}
\end{figure}

As seen in the aforementioned studies, DL-based channel estimation studies for ISAC generally utilized FNN, FCDNN, CNN, or ELM with supervised learning, which are data-driven methods. DL-based estimators are demonstrated to outperform linear estimators while maintaining a low computational complexity during inference. However, data-driven methods require a large number of training samples.

\begin{figure*}
    \centering
    \includegraphics[width=1\linewidth]{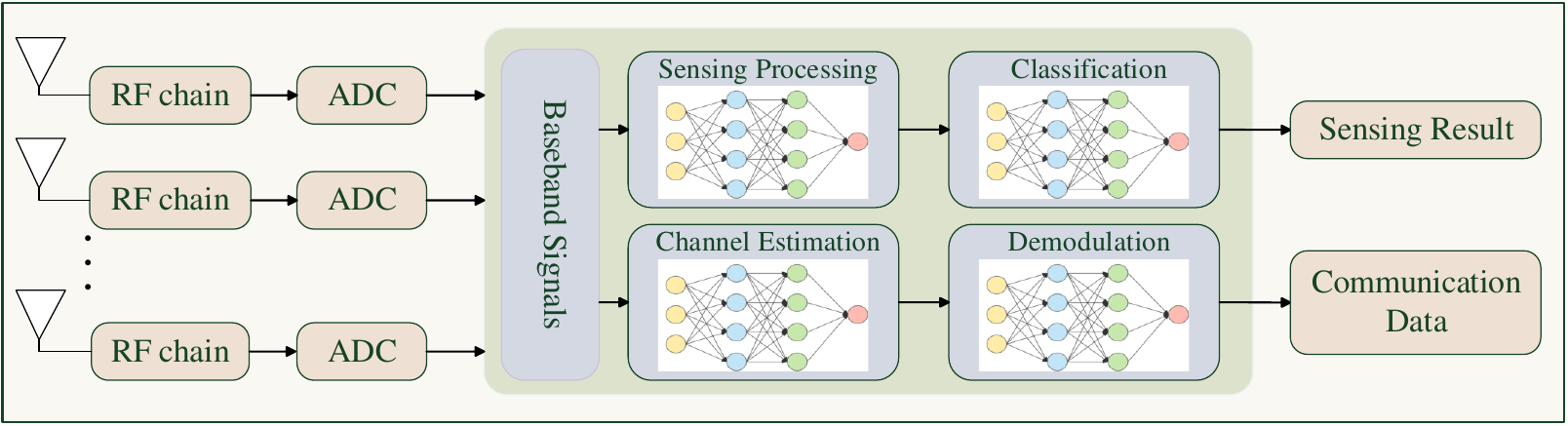}
    \caption{DL-based ISAC receiver architecture.}
    \label{fig:ISAC_receiver}
\end{figure*}

\begin{table*}[]

\centering
\caption{DL-based techniques for ISAC receivers.}
\footnotesize
\begin{tabular}{|l<{\centering}|m{1.25cm}<{\centering}|m{1.25cm}<{\centering}|m{2cm}<{\centering}|m{2.5cm}<{\centering}|m{4.5cm}<{\centering}|m{3.5cm}<{\centering}|}
\hline
\textbf{Ref.}  & \textbf{Learning
Strategy} & \textbf{Algorithm}    & \textbf{Input} & \textbf{Output} & \textbf{Pros}& \textbf{Notes}               \\ \hline
\cite{wuSensingIntegratedDFTSpread2023}            &Supervised Learning& FCDNN         & Received signals & \begin{itemize}[leftmargin=4pt]
    \item Demodulated communication data
    \item Target range and velocity estimations 
    \item Passive sensing parameters
\end{itemize}   & 
\begin{itemize}[leftmargin=4pt]
    \item ComNet: Better performance than MMSE under Doppler effects
    \item SensingNet: Higher accuracy for multi-target parameter estimation
    \item  Robust to Doppler effects,
phase noise and multi-path propagation
    \item Lower computational complexity
\end{itemize}
& 
\begin{itemize}[leftmargin=4pt]
    \item THz OFDM ISAC systems
    \item SensingNet: Active sensing parameter estimation
    \item ComNet: Simultaneous data symbol detection and passive sensing
\end{itemize}      \\ \hline
\cite{chenNeuromorphicIntegratedSensing2023a}      &Supervised Learning& SNN &  Received signals UWB signals & \begin{itemize}[leftmargin=4pt]
    \item Demodulated communication data
    \item Target detection
\end{itemize} &
\begin{itemize}[leftmargin=4pt]
    \item Neuromorphic ISAC
    \item Energy-efficient online decoding and sensing
\end{itemize} &  
\begin{itemize}[leftmargin=4pt]
    \item Neuromorphic computing for ISAC
    \item UWB signaling for ISAC
\end{itemize} \\ \hline

\cite{liuDeepLearningBased2022}                     & Supervised Learning&FCDNN and LSTM & Received communication signals under radar (FMCW) interference                    & Communication data symbols& 
\begin{itemize}[leftmargin=4pt]
    \item FCDNN has a higher accuracy when the symbol duration is short
    \item LSTM outperforms FCDNN when the symbol duration is long
    \item Achieves a better performance than the ZF
\end{itemize}  & 
\begin{itemize}[leftmargin=4pt]
    \item Performance of FCDNN is affected more by time-related distortion than LSTM
\end{itemize} \\ \hline

\cite{hu2024isac}                     &Supervised Learning & DNN Transformer & Received signals &
\begin{itemize}[leftmargin=4pt]
    \item Estimated communication data symbols
    \item Estimated target parameters: angle of arrival (AOA) and time delay (TD)
\end{itemize}
 & 
\begin{itemize}[leftmargin=4pt]
    \item Robust performance in signal detection with less training
\end{itemize} & 
\begin{itemize}[leftmargin=4pt]
    \item Exponentially increasing complexity with the number of antennas
\end{itemize}\\ \hline

\cite{jiang2024isac} & Supervised Learning&FCDNN & The received data signal, and the estimated CSI matrix,        & Demodulated communication data and passive sensing results & 
\begin{itemize}[leftmargin=4pt]
    \item Simultaneous passive sensing and demodulation
    \item Comparable sensing and communication performance with state-of-the-art algorithms.
\end{itemize}  & 
\begin{itemize}[leftmargin=4pt]
    \item Model-based DL-based passive sensing, signal detection, and channel reconstruction modules
module
\end{itemize} \\ \hline

\cite{gao5GNRHighPrecision2022} &Supervised Learning& CNN        & Channel frequency
response (CFR)        & Estimated {UE} positions                 & 
\begin{itemize}[leftmargin=4pt]
    \item Reduced training time compared to FCDNN
    \item Improved position estimation
\end{itemize} & 
\begin{itemize}[leftmargin=4pt]
    \item 5G NR indoor positioning
\end{itemize}\\ \hline
\cite{KoikeQNN2022}      &Supervised Learning& QNN        & Beam SNRs   & Human pose recognition      & 
\begin{itemize}[leftmargin=4pt]
    \item High accuracy, comparable to state-of-the-art DNN models.
    
\end{itemize} & 
\begin{itemize}[leftmargin=4pt]
    \item Proof-of-concept quantum neural networks
    \item Commercial WiFi devices are successfully employed for ISAC.
\end{itemize}\\ \hline

\cite{suarezDeepLearningaidedRobust2023}           &Supervised Learning & CNN        & Correlation matrix of the received signals       & Target delay and Doppler estimation& 
\begin{itemize}[leftmargin=4pt]
    \item Eliminate the need for threshold decisions
    \item Demonstrate robustness to changes in channel conditions
\end{itemize} & 
\begin{itemize}[leftmargin=4pt]
    \item OTFS signals are used for ISAC
    \item Target localization in the
delay-Doppler domain
\end{itemize} \\ \hline

\cite{liuVerticalFederatedEdge2022}                &Supervised Learning& CNN        & Sensing data matrix               & Objects/human motion recognition & \begin{itemize}[leftmargin=4pt]
    \item High objects/human motion recognition accuracy
    \item Low communication overhead due to exchanging only intermediate computed vectors.
\end{itemize}  &  
\begin{itemize}[leftmargin=4pt]
    \item Vertical federated edge learning
    \item Distributed sensing via {DL}
\end{itemize}   \\ \hline

\end{tabular}
\label{Tab: DL-receiver}
\end{table*}

\section{DL-based Receiver Techniques}

Conventional receiver algorithms generally employ linear techniques that have low complexity for demodulation, decoding, and signal processing. However, the sophisticated structure of ISAC waveforms may render such linear methods insufficient to reach the desired communication and sensing performance. On the other hand, non-linear and iterative algorithms tend to have high computational complexity and demand a significant amount of computational power and energy. Accordingly, DL-based techniques have recently been considered for the development of highly energy-efficient and practical receiver algorithms. DL-based  {techniques} have already been considered for various communication receiver operations, such as for spectrum sensing \cite{AhmedSensing2022}, channel equalization, and signal demodulation \cite{honkala2021deeprx, WangDLDemod2019}.

Developing an ISAC receiver typically demands more sophisticated algorithms and techniques than the transmitter techniques since hardware impairments, interference, synchronization (time, phase, and frequency) problems, or imperfect CSI estimation can easily degrade the performance of receiver algorithms and signal processing. Hence, the receiver techniques need to be robust to these problems. Fig.~\ref{fig:ISAC_receiver} illustrates an exemplary DL-based ISAC receiver architecture, where after obtaining the baseband signals via receiver RF chains and ADCs, the DL-based techniques are applied to the baseband signals for channel estimation, sensing processing, demodulation of the data, and radar target classification. Supervised learning is generally preferred in these studies since communication symbol estimation and sensing parameter estimation require foreknowledge of the communication symbols or parameters estimation.

Designing different architectures or combining multiple stages within a single DNN is also possible as a different architecture. Table~\ref{Tab: DL-receiver} summarizes the DL-based receiver ISAC studies, where DL-based techniques perform various blocks of the receiver signal processing, such as demodulating the communication data, estimating the target range and velocity, or target detection.

\begin{figure}
    \centering
    \includegraphics[width=1\linewidth]{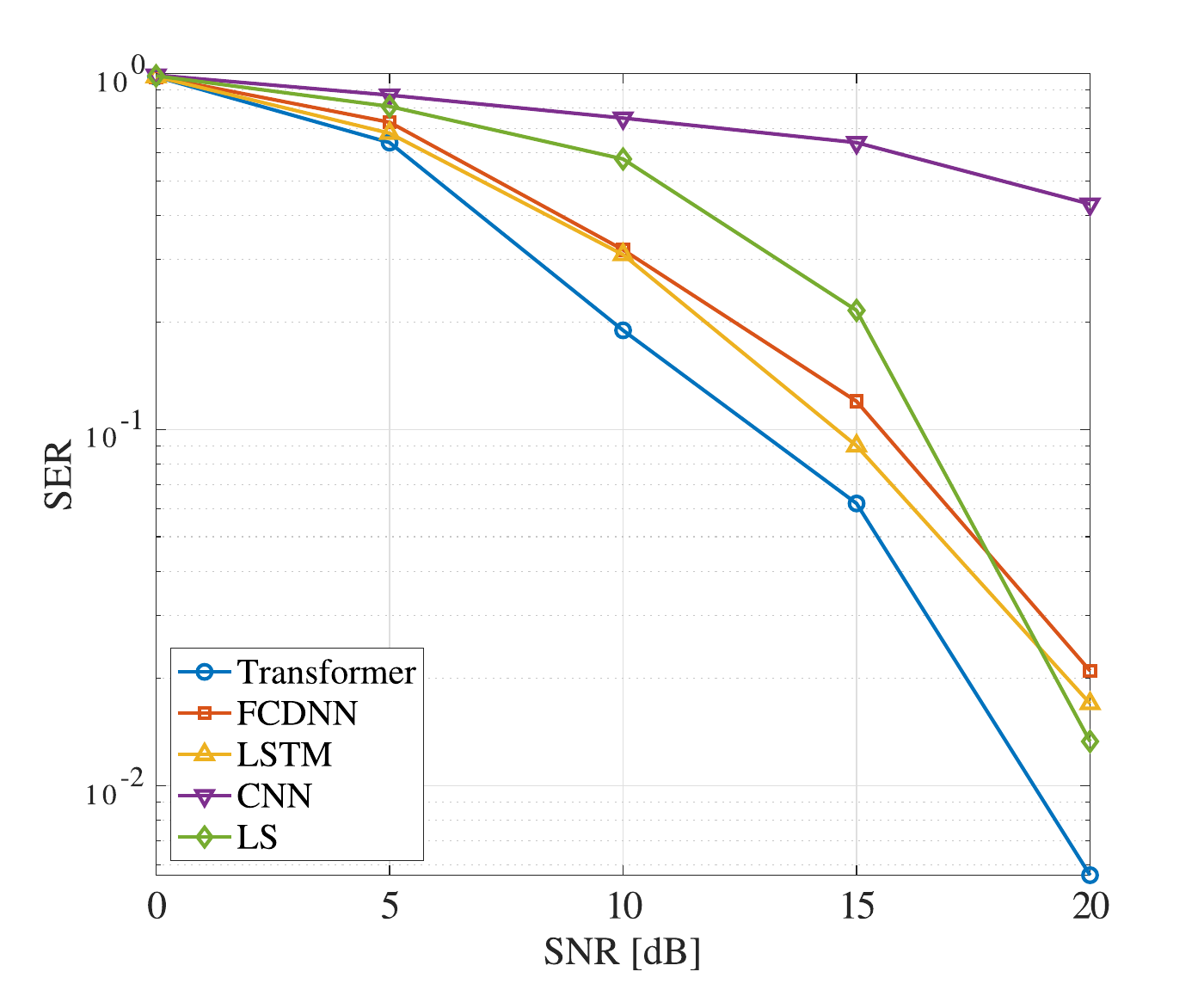}
    \caption{{Demodulation performance of various DL-based receivers for 64-QAM signals}.}
    \label{fig:data_demod}
\end{figure}

\subsection{Data Demodulation}

Hardware impairments, noise, interference, carrier signal phase or frequency offsets, or imperfect CSI can significantly degrade the demodulation performance of the receiver. To overcome these,  DL-based demodulation techniques have recently been proposed \cite{WangDLDemod2019, ZhangLSTM2020, AhmadDeepDeMod2022}, outperforming conventional demodulation techniques. For instance, researchers suggested a {transformer-based} method to demodulate QPSK signals at ISAC receivers in \cite{hu2024isac}, which is demonstrated to outperform other DL-based demodulators. {Fig~\ref{fig:data_demod} compares the performance of DL-based demodulators for the demodulation of 64-QAM symbols, where the transformer-based modulator proposed in \cite{hu2024isac} outperforms the FCDNN-based, LSTM-based, CNN-based, and LS demodulators when the same data set is used for training.  Moreover, it shows that the FCDNN-based and LSTM-based demodulators also outperform the LS demodulator, especially when the SNR is low.}

\begin{figure*}
    \centering
    \includegraphics[width=1\linewidth]{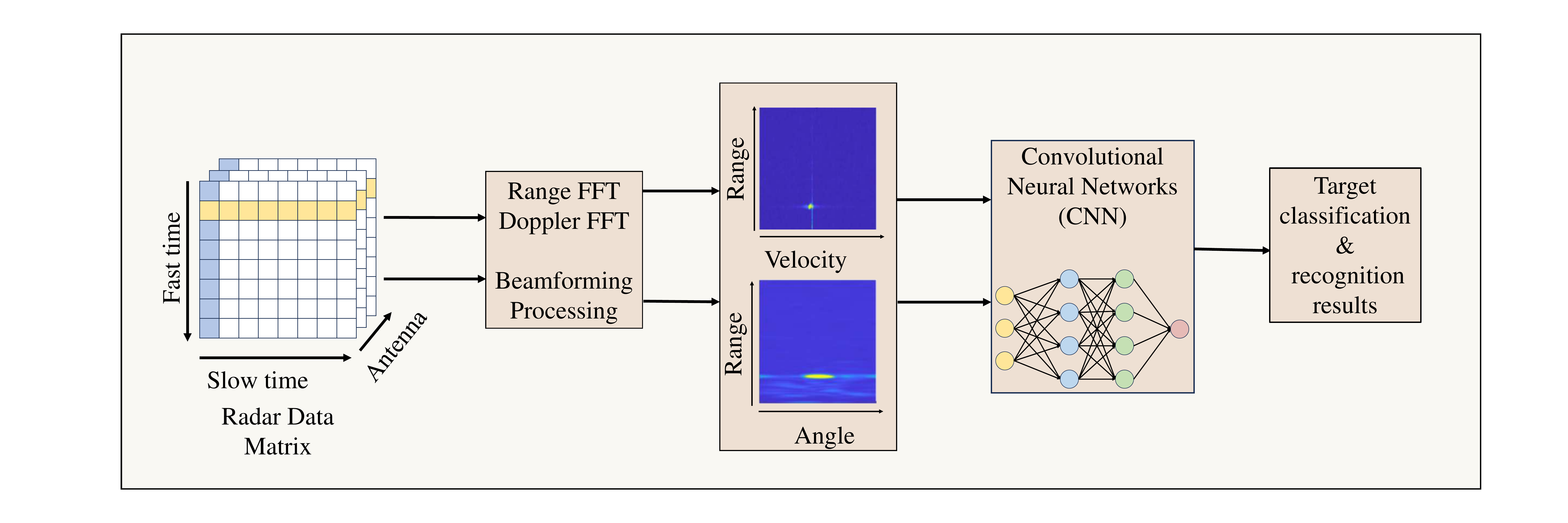}
    \caption{Radar target classification and recognition using CNNs.}
    \label{fig:target_recognition}
\end{figure*}

Another study presented in \cite{jiang2024isac} utilizes an FCDNN architecture to demodulate the communication symbols while simultaneously processing the received signals for sensing. On the other hand, a waveform for Terahertz (THz) ISAC systems is proposed in \cite{wuSensingIntegratedDFTSpread2023}, which is based on discrete Fourier transform spread orthogonal frequency division multiplexing (SI-DFT-s-OFDM). They also developed an ISAC receiver incorporating two neural networks that demodulate the communication data and estimate target range and velocity under hardware imperfections and non-linearities of THz systems, enhancing communication and sensing performance. Moreover, it alleviates the  Doppler effects and phase noise and can handle multi-target estimation. An FCDNN and LSTM-based architecture is proposed to improve the data demodulation performance under radar interference in \cite{liuDeepLearningBased2022}, which performs comparably to traditional detectors without requiring interference cancellation. Moreover, a neuromorphic computing solution based on a spiking neural network (SNN) for ultrawideband (UWB) impulse radio ISAC systems is proposed \cite{chenNeuromorphicIntegratedSensing2023a}, where received impulse signals are processed via an SNN for simultaneous demodulation of the communication data and target detection. Utilizing SNN instead of DNN provides a more energy-efficient solution for UWB ISAC systems.

\subsection{Radar Signal Processing}

Sensing processing in the ISAC receiver incorporates multiple stages, including receive beamforming, creating the sensing data matrix, target detection, target parameter estimation, and target classification and recognition, similar to traditional radar systems. DL-based architecture is already applied in multiple stages of traditional radar systems, such as target detection, parameter estimation,  and target classification and recognition.

DL-based algorithms for radar signal processing have been increasingly studied in recent years, especially for radar waveform and array design,  radar waveform recognition, target recognition and classification, and also suppression of jamming, clutter, and interference  \cite{GengRadar2021, SARTargetRecognition2018_XHJ, TargetRecognition2021_XHJ, TargetRecognition2023_XHJ, RadarClassification_ANN_2021_XHJ, RadarClassification_CNN_2020_XHJ, RadarClassification_CNN_2021_XHJ, RadarDataProcessing_DNN_2020_XHJ, ToppleTargetRec2021}. Some of these methods utilize supervised learning, such as \cite{SARTargetRecognition2018_XHJ, TargetRecognition2023_XHJ}, to classify or recognize the targets in synthetic-aperture radar images. Moreover,  self-supervised learning, which is a subset of unsupervised learning, is also considered for target recognition \cite{MuzeauSelfSupervisedRadar2022}. Self-supervised models learn the representative characteristics from unlabeled
data themselves without requiring labeled data. Furthermore, transfer learning is also employed for the target recognition and classification \cite{RadarClassification_CNN_2021_XHJ, RadarClassification_ANN_2021_XHJ}. In general, CNN-based algorithms are used for radar receiver techniques since they can be efficiently applied to radar images, e.g., range-velocity and angle-range images. These studies mentioned above considered only radar systems. Radar signal processing in ISAC systems is reviewed below under two categories: target detection and parameter estimation, and human gesture and motion recognition.

\subsection{Target Detection and Parameter Estimation}
Range, velocity, and angle are the main target parameters that are aimed to be estimated in radar systems after detecting targets. Estimating these parameters in highly sophisticated environments and with low SNRs becomes challenging. Consequently, DL-based  {techniques} have already been widely studied to estimate such parameters in ISAC systems. Supervised learning with FCDNN is utilized for passive and active sensing and target parameter estimation in THz systems \cite{wuSensingIntegratedDFTSpread2023}, where the radar processing delivered high accuracy while having a lower computational complexity than MUSIC. In another study, UWB signals are processed via spiking neural networks, i.e., neuromorphic computing, to perform target detection in a UWB impulse radio system \cite{chenNeuromorphicIntegratedSensing2023a}. Moreover, a DNN transformer architecture is employed to jointly estimate the communication data symbols and the range and angle of the targets, providing a robust estimation performance \cite{hu2024isac}. Model-based DL is also considered to jointly perform demodulation and passive sensing \cite{jiang2024isac}, which reduces the training time and improves the sensing performance. Moreover, target delay and Doppler estimation for OTFS signals are also studied using CNN in \cite{suarezDeepLearningaidedRobust2023}, where the proposed DL-based target parameter estimator can estimate the target parameters without using a threshold, and it is robust to variations of the channel. The aforementioned studies mainly employ the supervised learning approach to estimate the target parameters, while CNN, FCDNN, and Transformers are employed as the architectures.

\subsection{Human Gesture and Motion Recognition}

Human gesture and motion recognition can be performed via active or passive sensing. Moreover, ISAC signaling can leverage human gesture and motion recognition. For instance, indoor ISAC systems can recognize human motions when communicating with indoor devices to provide sensing data for security, health care, or other applications.  A vertical federated edge learning {technique} is proposed to recognize objects or human motions by utilizing distributed ISAC signals in \cite{liuVerticalFederatedEdge2022}, where distributed edge devices employ ISAC signals for sensing and share intermediate computed vectors for collaborative sensing. This method reaches around 98\% recognition accuracy, resulting in 8\% accuracy enhancement compared to the benchmark methods. Moreover, a proof-of-concept study has proposed a quantum neural network (QNN) architecture for human pose recognition, which achieves a similar estimation accuracy compared to state-of-the-art methods\cite{KoikeQNN2022}. Further studies on human gesture and motion recognition powered by ISAC can improve security, privacy, and healthcare applications.

\begin{table}
\centering
{\caption{Real multiplications for each layer of different NN architectures \cite{FreireComputationalComplexity2024}.}
\begin{tabular}{|c|c|}
\hline
\textbf{Network type} & \textbf{Real multiplications} \\
\hline
MLP & $N_N N_I$ \\
\hline
1D-CNN & $N_F N_I N_K N_O$ \\
\hline
RNN & $N_S N_H (N_I + N_H)$ \\
\hline
LSTM & $N_S N_H (4N_I + 4N_H + 3)$ \\
\hline
\end{tabular}
\label{tab:real_multiplications}}
\end{table}

\begin{table*}[]
\centering
{\caption{Comparison of optimization-based, model-based, learning-based, and data-driven learning-based methods.}
\small
\begin{tabular}{|l|l|l|l|}
\hline
                        & \textbf{Optimization-based} & \textbf{Model-based DL} & \textbf{Data-driven DL} \\ \hline
\textbf{Training Data}  & -                         & Less Data                      & Data-hungry                   \\ \hline
\textbf{Training Time}  & -                         & Short                        & Long                        \\ \hline
\textbf{Execution Time} & Long (0.1 $-$ 30 seconds)   &  Short (microseconds $-$ milliseconds)      &  Short (microseconds $-$ milliseconds)     \\ \hline
\end{tabular}
\label{tab:complexity_com}}
\end{table*}

{\section{Computational Complexity of DL-based Techniques}}
{One of the major advantages of utilizing DL-based techniques is to deliver near-optimum results with low computational complexity compared to iterative or optimization-based methods. Although their training may take time due to requiring a large amount of data, their execution time can be in microseconds or milliseconds, enabling their implementation in real-time systems. This section highlights the computational complexity reduction offered by DL-based  {techniques} in comparison with iterative or optimization-based methods.}

{Table~\ref{tab:real_multiplications} compares a computational complexity metric, namely real multiplications for each layer, for widely used NN architectures, where  $N_N$, $N_I$, $N_F$, $N_K$, $N_O$, $N_S$, $N_H$, $N_I$ denote the number of neurons, number of features in the input vector, number of filters, kernel size,
output size, input kernel time sequence size, and number of hidden units, respectively \cite{FreireComputationalComplexity2024}. It can be seen that the computational complexities of DL-based  {techniques} are mainly related to their architectures, hence, it is difficult to provide a direct computational complexity comparison between the DL-based  {techniques} and optimization-based or iterative methods. Therefore, many studies compare their complexities in terms of execution times. Table~\ref{tab:complexity_com} presents a comparison of optimization-based, model-based, and data-driven methods in terms of execution times and performance. The details of this comparison are explained in the training complexity and inference complexity subsections below.} 

\subsection{{Training Complexity}}

{The training of DL-based  {techniques} generally requires a large amount of data and long training times. The training times of learning-based methods are mainly related to NN architecture, input and output sizes, and training data set size. However, various techniques have also been proposed to reduce the training time of DL-based  {techniques}. For instance, for the same amount of data, an extreme learning machine, which is a single-layer NN architecture and does not employ gradient during training, can be trained 5 times faster than FCDNN, while achieving a similar channel estimation performance with FCDNN for channel estimation \cite{LiuEL_CSIA2023}. On the other hand, the amount of data required and training time can be substantially reduced by model-based, e.g., deep-unfolding, methods since these methods utilize domain knowledge, leading to more computationally efficient training with less data \cite{ZhangLowComplexity2025}. Thus, carefully designing a problem-specific NN architecture that incorporates domain knowledge can reduce training times and the amount of data required for training.}

\subsection{{Inference Complexity}}
{The DL-based  {techniques} can be swiftly executed since only matrix operations need to be performed during inference. This greatly reduces the execution time compared to optimization-based methods. For instance, two iterative methods were proposed for symbol-level ISAC precoder design, one of which employs penalty dual decomposition, majorization-minimization, and block coordinate descent methods to iteratively design the precoder, while the other one employs the augmented Lagrangian method with Riemannian Broyden-Fletcher-Goldfarb Shanno algorithm. The execution times of these algorithms are 9.91 seconds and 0.233 seconds, respectively \cite{PDD2021}. To reduce the execution time, the DL-based symbol-level ISAC precoding methods are proposed \cite{JiangSLP2025}, which can design the sub-optimum  ISAC waveforms within 0.01-0.02 seconds. Note that the iterative algorithms mentioned above outperform DL-based  {techniques} in terms of communication sum rate, however, DL-based  {techniques} offer more practical implementation compared to iterative methods since real-time communication systems may need to swiftly generate a precoder every few hundred milliseconds due to varying channel conditions. It is observed that the execution time of iterative methods can take up to 10 s or more, depending on the number of antennas and {UEs}, while corresponding DL-based  {techniques} can be executed in a few hundred milliseconds, in most of the studies \cite{QiDeepLearning2024, ZhangLowComplexity2025, ZhangJointDesign2025, JiangSLP2025, PDD2021, wuSensingIntegratedDFTSpread2023}.  Moreover, the execution time of iterative or optimization-based methods may exponentially increase with the number of antennas, while the execution time of DL-based  {techniques} slightly increases with the number of antennas, since DL-based  {techniques} mainly perform matrix and vector multiplications that can be efficiently executed in parallel \cite{ZhangLowComplexity2025, ZhangJointDesign2025}.} 

{It is worth noting that DNN architecture and its size also determine the inference time. However, DNN models can also be further optimized to reduce their computational complexity, execution time, and memory usage via weight quantization, pruning, and knowledge distillation \cite{SzeEfficientDNN2017}. For instance, a DL-based ISAC precoder with 6-bit quantized and 10\% pruned DNN is shown to achieve 96.1\% of the sum rate achieved by the complete DNN model. At the same time, its memory usage and computational requirements are only 16.88\% of the complete DNN model \cite{TemizUnsupervised2025}.}


{In addition to the computational advantages presented above, DL-based  {techniques} can run on specifically developed low-power NN accelerators or GPU-based hardware that can execute them swiftly with high energy efficiency\cite{9739030}. Such low-power NN accelerators can achieve very high energy efficiency, e.g., around 25TFLOPS/W, while consuming only a few hundred milliwatts of power \cite{LeeAutomotive2023, LatotzkeDNNHardware2021}. This makes DL-based  {techniques} more practical compared to optimization-based methods, even for low-power edge and IoT devices \cite{LeeHardware2020}. Consequently, DL-based  {techniques} can be used in base stations and {UEs} to develop ISAC systems without causing a significant computational burden and power consumption.}

\section{Challenges and Future Research Directions}
{The fusion of DL with signal processing techniques for joint sensing and communications holds significant promise for future networks.} ISAC can potentially transform various systems and technologies, including autonomous vehicles and smart cities. Even though DL-based techniques can remarkably improve the performance and energy efficiency of ISAC systems, significant challenges still need to be addressed. These challenges also pave the way for opportunities and future research directions. In this section, we present the challenges encountered in DL-based ISAC systems and possible related future research directions. 

\subsection{Challenges for DL-based Techniques for ISAC}

{While DL-based techniques may significantly enhance the performance of ISAC systems and enable their implementation, several challenges must be addressed to fully realize their potential in real-time and sophisticated networks, as:}

\begin{itemize}
    \item {\textbf{Data Availability and Reliability:} To train DL models via supervised or unsupervised learning approaches, a large amount of data representing the realistic ISAC scenarios and channels needs to be collected in advance. This data can be obtained through measurements \cite{XuExperimental2022, TemizISACSensing2023}, or realistic system simulations, such as using computationally expensive ray tracing-based simulations\cite{LiuISACChannel2024}. However, obtaining data for all possible ISAC scenarios can be extremely time-consuming or expensive.} 
    \item {\textbf{Model Complexity:} Although DL models excel at producing near-optimal solutions for various tasks, their complexity can be prohibitive, especially when a large number of layers and nodes are involved or sophisticated architectures like LSTM and transformers are employed in the model. This significantly increases the computational complexity of training, requiring highly powerful and costly computational resources \cite{FournierTransformers2023, hu2021model}.}
    \item {\textbf{Interpretability and Predictability:} Data-driven DL models operate as black boxes, hence their decision-making processes remain largely opaque and are not yet fully understood \cite{chakraborty2017interpretability}. Furthermore, their behavior in unseen scenarios is unpredictable, rendering them unsuitable for critical applications where reliability and transparency are necessary.}
    \item {\textbf{Scalability and Generalization:} DL models are typically trained for a specific number of inputs and outputs using a dataset consisting of particular cases. Thus, these models are optimized for scenarios that closely resemble the training data, which may limit their generalization to unforeseen situations. However, DL-based techniques for ISAC systems must be flexible and scalable to handle swiftly varying parameters such as the number of UEs, the trade-off between the sensing and communications, interference between the sensing and communication signals, and clutter conditions \cite{TemizUnsupervised2025}. Moreover, DL models need to be scalable and flexible in terms of resource scheduling, data management, parallelization, and training \cite{mayer2020scalable}.}
     \item {\textbf{Hardware Limitations:} DL models need to operate in real-time, hence their computational complexity must be low to operate on limited computational and memory resources \cite{chen2020deep}. Especially, techniques for UEs must have low computational complexity and power consumption due to the hardware limitations and energy constraints of commonly used UEs, such as smartphones and IoT devices. \cite{li2018learning}.}
     \item {\textbf{Networking and Cooperation:} Networking of multiple ISAC cells is necessary to improve both sensing performance and communication coverage. Thus, the DL-based techniques developed for ISAC need to be able to work in cooperation with each other and share data between models in a network. Networking is also necessary to reduce the interference between the communication and sensing tasks of base stations. Moreover, this networking and cooperation need to be performed securely to maximize the sensing privacy and communication data security. Especially {FL}-based approaches can be utilized to enable secure networking and cooperation\cite{jiang2025federated, liuVerticalFederatedEdge2022}.  } 
\end{itemize}
{The challenges presented above need to be addressed before the deployment of DL-based techniques in ISAC networks. Thus, there exist significant open research opportunities, which are categorized and discussed below.}

\subsection{Model-based Deep Learning for ISAC}

One of the challenges encountered is the black-box structure of {data-driven} DL-based  {techniques}, hence, they are not interpretable, and they may produce unpredictable results in some situations \cite{li2022interpretable}. To overcome this issue, communication and signal processing theory and domain knowledge can be utilized within {DL} techniques to design model-based {DL} approaches for {ISAC.} {Model-based DL methods incorporate prior knowledge or theories about the system into the learning process, hence leading to more explainable AI models \cite{zappone2019wireless, shlezinger2023model}.} Model-based DL, therefore, leads to more efficient, accurate, and understandable models for ISAC applications \cite{mateos2023model, PulkinenISAC2024}. 

Model-based { DL methods} are designed to replicate the steps of signal processing algorithms by neural network layers to learn and predict various system parameters of ISAC systems or {optimize} their performance by intelligently adapting to changing environments or requirements. Developing model-based DL techniques requires solid domain knowledge in addition to knowledge of {DL} {architectures.} Model-based {DL} can be developed for numerous modules of ISAC systems, such as waveform design, channel estimation, and {communication and sensing data processing \cite{PulkinenISAC2024, ZhangJointDesign2025}.}

\subsection{Scalable and Distributed Deep Learning Models}
One of the fundamental problems with using {DL} in ISAC systems is the limited scalability of the {DL} models \cite{XuScalable2020}. {DL} models are generally trained to operate with a fixed number of inputs and outputs. For instance, changing the number of antennas or {UEs} may require re-training the model if it is developed based on a conventional {DL} model. Hence, more studies are necessary to address the scalability of DL-based techniques for ISAC systems. For instance, model-based DL techniques can be one of the solutions for scalability issues. 

On the other hand, distributed learning techniques, especially {FL}, can be applied in ISAC systems to enhance their scalability at a network level. {FL} will enable training large models without sharing enormous amounts of data and causing privacy concerns \cite{liuVerticalFederatedEdge2022}. Multi-cell ISAC networks or IoT networks will significantly benefit from distributed learning approaches since each base station and the {UE} will also facilitate the scalability of the entire network.

\subsection{Lightweight Deep Learning Models}

Although training {DL} models may require high computational resources, inference of trained {DL} models can be performed with much less computational complexity. Hence, they can be run even on resource-limited low-power devices such as mobile devices or IoT nodes \cite{chen2020deep, li2018learning}.
Moreover, the computational resource requirement of DL models can be further reduced by various methods such as node pruning, weight pruning, quantization, and knowledge distillation without significantly compromising their performance. \cite{wang2022lightweight}. Consequently, it is possible to develop lightweight {DL} models. {For instance, quantization and pruning are used to reduce the computational complexity, memory usage, and improve the energy efficiency of DL-based precoder for ISAC systems \cite{TemizUnsupervised2025}.}

Lightweight {DL} models are designed to be efficient and compact, making them suitable for environments with limited computational resources. These models are particularly useful in applications operating on resource-constrained devices such as ash mobile devices, embedded systems, or IoT devices, where computing power, memory, and energy are constrained \cite{wang2022lightweight}. The goal of lightweight models is to maintain a balance between model size, speed, and accuracy. Lightweight DL models will also reduce the hardware cost and increase the energy efficiency.

\subsection{Interference Cancellation}

The received signals may include interference from various sources, including self-interference, inter-channel interference, interference from sensing signals, and other devices. For instance, inter-user interference, inter-cell interference, self-interference, or interference between communication and radar systems are commonly observed \cite{liu2023distributed, zheng2019radar, chiriyath2017radar} in communication systems. Accordingly, ISAC receivers need to perform interference cancellation to enhance the performance of demodulation and sensing, especially in congested RF environments. {DL} algorithms can be employed to predict and alleviate interference patterns. AI-driven adaptive algorithms are then used to dynamically adjust system parameters in real-time based on the interference environment. DL-based interference cancellation techniques are proposed for MIMO and non-orthogonal multiple access (NOMA) communication systems \cite{sim2020deep, van2022deep, guo2019dsic, shlezinger2020deepsic}. Interference cancellation at the receiver of ISAC systems is more complicated than communication or sensing systems. Thus, DL-based techniques need to be explored for interference cancellation in ISAC systems.

\subsection{Target Recognition and Classification}

DL-based architectures are superior in target classification and recognition compared to conventional methods \cite{RadarClassification_ANN_2021_XHJ, RadarClassification_CNN_2020_XHJ, TargetRecognition2023_XHJ}. CNN-based methods are widely used for target recognition and classification since they require sensing images obtained via ISAC or radar systems. 
 
Target recognition and classification have been explored in radar systems; however, they have not yet been applied in the ISAC systems. CNN-based methods can also be used with ISAC sensing images, and they can be fine-tuned to {utilize the features} of ISAC signals instead of pure radar signals to improve {target parameter estimation, recognition, and classification \cite{MateosSemi2024, suarezDeepLearningaidedRobust2023}}.

\subsection{Synchronization}

In communication systems, achieving synchronization at the receiver end is crucial for accurately decoding signals. This process involves three key components: time, frequency, and phase synchronization. Time synchronization ensures that the receiver correctly identifies each symbol or data packet's start and end, aligning its internal clock with the transmitter's timing. Frequency synchronization adjusts the receiver's frequency to match the transmitter's, compensating for any drift or discrepancy that might occur due to Doppler shifts or oscillator inaccuracies. Phase synchronization is necessary to maintain the integrity of the signal's phase information, which is critical for decoding complex modulation schemes. Collectively, these synchronization mechanisms enable the receiver to accurately reconstruct the transmitted data, ensuring efficient and reliable communication. 

DL-based techniques have already been applied for the synchronization in communication systems \cite{aoudia2022deep, zhang2023distributed, abakasanga2023unsupervised, wu2019deep}. Synchronization in ISAC systems is more challenging than only communication systems due to {the requirement for} synchronization for both communication and sensing, especially in the case of distributed ISAC systems. In a distributed ISAC system, the synchronization between the ISAC base station and {UEs} may require a substantial amount of communication overhead. Thus, DL-based techniques can be developed to perform real-time synchronization in ISAC systems with a limited amount of communication overhead.

\subsection{Networking and Data Fusion}

Networking of communication systems, i.e., multi-cell communication, enhances the network's coverage and enables better interference and spectrum management. On the other hand, networking radar sensors or sensing data fusion enables target localization, improves sensing performance and coverage, and provides robustness against jamming \cite{TemizRadarNetwork2022, DhulashiaJamming2022, JingFusiaonIoT20203}. Thus, networking is necessary for both communication and sensing systems. ISAC systems need to be networked to provide a communication infrastructure within a larger area, reduce interference, fuse the sensing data to deliver more accurate sensing, and enable more efficient utilization of resources \cite{jiang2025federated}.

The networking of ISAC systems is a relatively recent topic; hence, it has numerous research opportunities. For instance, interference management is a complex problem since various sources of ISAC transmitters can interfere with each other. Coordinated control of closely operating transmitters is required to minimize the interference; however, reaching an optimum strategy can be challenging due to having a large number of transmitters or parameters. {DL} or {RL}-based coordination can handle this complexity, enabling cooperative sensing and mitigating interference among multiple ISAC cells \cite{Lu2024Deep}.  {ISAC network architecture can be designed to utilize {FL} to maximize the both sensing and communication performances \cite{jiang2025federated, liu2023ai}, improve the privacy of the sensing data \cite{HuFLISAC2025, liuVerticalFederatedEdge2022}, reduce the communication overheads \cite{OuyangCommFL2024}, and resource allocation \cite{LiuMultiTaskRA2024} in ISAC networks.} 

\subsection{{Integrated Sensing and Semantic Communications}}

Semantic communication is a new paradigm that transmits semantic information (meanings) rather than bits, hence substantially increasing the amount of information transmitted. Moreover, semantic communication also improves data security since the semantic information needs to be extracted from the received bits. The interpretation of the received bits and their transformation into semantic information can be performed by DL-based  {techniques}, such as {transformers and} large language models (LLMs) \cite{LuoSemantic2022, JiangLLM2024}. 

Semantic communication can be considered within ISAC systems towards 6G and beyond communication networks to improve information transmission rates and data security \cite{Sagduyu6GSemanticISAC2024}. {A transformer-based method is proposed to estimate the semantic communication and sensing channel while reducing the piloting overhead \cite{zhang2024ai}. Another study proposed a semantic-based CSI feedback method that utilizes autoencoders to reduce the feedback overhead \cite{ZhuSemantic2025}. Moreover, a DRL-based semantic-aware resource allocation method is proposed for vision-assisted ISAC systems \cite{LuSemantic2024}.}  
Semantic communication can be considered within ISAC systems towards 6G and beyond communication networks to improve information transmission rates and data security \cite{Sagduyu6GSemanticISAC2024}. Moreover, the acquired sensing information, along with communications in ISAC, can also enhance the performance of semantic communication by providing information about the channel and environment to the ISAC nodes.

\subsection{Realistic System Models and Simulations}
Most ISAC scenarios use some simplified assumptions due to the high complexity of real scenarios. However, implementing DL-based  {techniques} will allow researchers to study more realistic scenarios. Modeling such scenarios is necessary to understand the practical benefits of ISAC systems and related research problems. For instance, the SDP3 simulation framework, which is based on {DL}, is proposed to model an ISAC system and predict the performance and trade-offs of {ISAC} systems under realistic scenarios.  \cite{LiSDP32023}. 

The ISAC systems can also be studied and optimized for various realistic applications, such as autonomous driving, where an ISAC system can effectively provide wireless communication and sensing functions, reducing the amount of hardware, cost, and energy consumption. \cite{Dataprocessing2022}. Such systems can be modeled in a more realistic way to evaluate the practical benefits of ISAC.

\subsection{Prototype Development}

The real-time implementation of {DL} techniques requires specific hardware for energy and hardware-efficient operations instead of using general-purpose CPU architectures. Thus, many hardware platforms have already been developed for {DL} implementation. Even small IoT devices can now have {DL} accelerators that significantly improve energy efficiency. A recent survey of ML and {DL} acceleration platforms shows that a wide range of design techniques are used to design hardware and energy-efficient {DL} acceleration platforms, such as neural approximation, in-memory computing, and multi-stage acceleration\cite{9739030}. Such hardware platforms are developed with various hardware architectures such as FPGA \cite{9447019}, ASIC, in-memory \cite{9739030}, and GPU \cite{S2542660522001469}. Low-cost and low-power architectures are essential, especially for IoT and edge computing devices that operate on batteries \cite{9806150, 9447019}.

Prototype development is also an essential stage in proving the performance of DL-based ISAC techniques. It involves building an initial version of the ISAC system to show its features, usability, and overall functionality. Moreover, measurements in the lab and field trials can provide insight into their performance in real-world applications. Such prototypes can include various DL-based algorithms for ISAC transceivers to generate the ISAC signals, demodulate and decode the communication data, process the sensing data, and recognize the target. Such prototype development can be performed using various available DL accelerators and RF hardware. For instance, Jetson Nano or FPGAs can be used for efficient and accelerated {DL} inference \cite{gozuoglu2024cnn}. On the other hand, RFSoC \cite{peters2022arestor, peters2023modular} and software-defined radio (SDR) platforms \cite{banerjee2022deep, chung2022leveraging} can be utilized for {real-time} analog and RF signal processing.

\section{Conclusion}
This article has reviewed the DL-based techniques for ISAC design and optimization and summarized future directions and research opportunities in this area. DL-based techniques can enable the implementation of ISAC systems in 6G and beyond communication networks by reducing the amount of hardware used, computational complexity, and energy consumption. The powerful data-driven and model-based DL techniques offer tremendous opportunities for waveform design, channel estimation, and receiver design for ISAC systems, as seen in the state-of-the-art studies reviewed in this survey.} A wide range of DL methods, such as {FCDNN, CNN, LSTM, FL, RL, and transformers} can be employed. These DL architectures can be trained by supervised learning, unsupervised learning, or semi-supervised learning strategies over distributed hardware to realize near-optimum techniques for ISAC systems that can operate in real-time. On the other hand, the scalability and flexibility of DL-based algorithms, the ability to generate sufficient data for training, the implementation of these methods on mobile devices, and network-level ISAC systems need to be explored further to effectively employ them in future generation wireless networks.

\section*{Acknowledgment}
This work was supported in parts by the NSFC Grants 62361136811, 62174091, 62201294, TUBITAK Grant 123N800, METU BAP Grant AGEP-301-2025-11558, and EPSRC Grant EP/S028455/1.

\bibliographystyle{IEEEtran}
\bibliography{ref.bib}
\end{document}